%% file: sample.tex
\documentclass[sigconf]{aamas}

\usepackage{balance} %

\usepackage{caption}
\usepackage{mathtools}
\usepackage{amsmath}
\usepackage{tikz}
\usepackage{tikz-3dplot}
\usepackage{pgfplots}
\usepgfplotslibrary{ternary}
\usetikzlibrary{calc,shapes}
\usepackage{todonotes}
\usepackage{dsfont}
\usepackage{enumitem}
\usepackage{tikz}
\usepackage{amsthm}
\usepackage{thm-restate}
\usepackage{booktabs}
\usepackage{multirow}
\usepackage{newfloat}
\usepackage{wrapfig}
\usepackage{dsfont}
\usepackage{comment}
\usepackage{multirow}
\usepackage{mleftright}
\usepackage[capitalise,noabbrev]{cleveref}
\usepackage{xcolor-material}
\usepackage{listings}
\usepackage{subcaption}
\usepackage{blindtext}

\newtheorem{example}{Example}

\newtheorem{definition}{Definition}

\newcommand{\eg}{\emph{e.g.}}
\newcommand{\ie}{\emph{i.e.}}

\newcommand{\pl}{\mathcal{P}}
\newcommand{\Ss}{\mathcal{S}}

\newcommand{\chance}{\textsc{C}}

\newcommand{\aspace}{\mathcal{A}}
\newcommand{\Z}{\mathcal{Z}}
\newcommand{\buffer}{\mathcal{M}}
\newcommand{\game}{\mathcal{G}}
\newcommand{\pref}{\game^\ast}

\DeclareMathOperator*{\argmax}{arg\,max}

\definecolor{col1}{HTML}{2f6c9e}
\definecolor{col2}{HTML}{d16800}

\newcommand{\tone}{\textsf{T1}}
\newcommand{\ttwo}{\textsf{T2}}
\newcommand{\team}{\textsf{T}}
\newcommand{\opp}{\textsf{O}}
\newcommand{\lefta}{\textsf{L}}
\newcommand{\righta}{\textsf{R}}
\newcommand{\aloss}{\text{\textsc{A}-loss}}

\newcommand{\SIMSQMIX}{\text{SIMS-\textit{i}QMIX}}
\newcommand{\SIMSNFSP}{\text{SIMS-\textit{i}NFSP}}

\newcommand*\circled[1]{\tikz[baseline=(char.base)]{
		\node[shape=circle,draw,inner sep=1pt] (char) {\scriptsize #1};}}

\setcopyright{ifaamas}
\acmConference[AAMAS '21]{Proc.\@ of the 20th International Conference on Autonomous Agents and Multiagent Systems (AAMAS 2021)}{May 3--7, 2021}{Online}{U.~Endriss, A.~Now\'{e}, F.~Dignum, A.~Lomuscio (eds.)}
\copyrightyear{2021}
\acmYear{2021}
\acmDOI{}
\acmPrice{}
\acmISBN{}

\acmSubmissionID{158}

\title[]{Multi-Agent Coordination in Adversarial Environments through Signal Mediated Strategies}

\author{Federico Cacciamani}
\affiliation{
 \institution{Politecnico di Milano}
}
\email{federico.cacciamani@polimi.it}

\author{Andrea Celli}
\affiliation{
 \institution{Politecnico di Milano}
}
\email{andrea.celli@polimi.it}

\author{Marco Ciccone}
\affiliation{
 \institution{Politecnico di Milano}
}
\email{marco.ciccone@polimi.it}

\author{Nicola Gatti}
\affiliation{
 \institution{Politecnico di Milano}
}
\email{nicola.gatti@polimi.it}

\begin{abstract} 
Many real-world scenarios involve teams of agents that have to coordinate their actions to reach a shared goal. We focus on the setting in which a team of agents faces an opponent in a zero-sum, imperfect-information game. Team members can coordinate their strategies before the beginning of the game, but are unable to communicate during the playing phase of the game. This is the case, for example, in Bridge, collusion in poker, and collusion in bidding. 
In this setting, model-free RL methods are oftentimes unable to capture coordination because agents' policies are executed in a decentralized fashion. Our first contribution is a game-theoretic centralized training regimen to effectively perform trajectory sampling so as to foster team coordination. When team members can observe each other actions, we show that this approach provably yields equilibrium strategies. Then, we introduce a signaling-based framework to represent team coordinated strategies given a buffer of past experiences. Each team member's policy is parametrized as a neural network whose output is conditioned on a suitable exogenous signal, drawn from a learned probability distribution. By combining these two elements, we empirically show convergence to coordinated equilibria in cases where previous state-of-the-art multi-agent RL algorithms did not. 
\end{abstract}


\newcommand{\BibTeX}{\rm B\kern-.05em{\sc i\kern-.025em b}\kern-.08em\TeX}

\begin{document}

\pagestyle{fancy}
\fancyhead{}

\maketitle

\input{text/1.intro}
\input{text/2.related_work}

\input{text/3.preliminaries}
\input{text/4.coordination}
\input{text/5.centralized}
\input{text/6.sims}
\input{text/7.experiments_v2}
\input{text/8.conclusions}

\input{supp_content}

\bibliographystyle{ACM-Reference-Format} 
\bibliography{biblio}

\end{document}

%% file: text/1.intro.tex
\section{Introduction}
\label{sec:intro}

In many strategic interactions agents have similar goals and have incentives to team up, and share their final reward. In these settings, coordination between team members plays a crucial role. We focus on \emph{ex ante coordination}, where team members have an opportunity to discuss and agree on tactics before the game starts, but will be unable to communicate during the game, except through their publicly-observed actions. Consider, as an illustration, a poker game where some players are colluding against some identified target players and will share the final winnings after the game. Another instance of this problem is the card-playing phase of Bridge, in which two {\em defenders} have to coordinate their actions against the {\em declarer}, but they are prohibited from communicating by the rules of the game.

Finding an optimal equilibrium with ex ante coordination is \textsf{NP}-hard and inapproximable~\cite{celli2018computational}. \citet{celli2018computational} introduced the first algorithm to compute optimal coordinated strategies for a team playing against an adversary. At its core, it is a column-generation algorithm exploiting a hybrid representation of the game, where team members play joint normal-form actions while the adversary employs sequence-form strategies~\cite{koller1996efficient}.
It is crucial to observe that the number of joint normal-form actions of the team grows exponentially in the size of the game tree, which makes them impractical when dealing with games of medium/large size.
More recently, \citet{Farina&Celli2017} proposed a variation of the Fictitious Play algorithm, namely Fictitious Team-Play (FTP), to compute an approximate solution to the problem. Both approaches require to iteratively solve Mixed-Integer Linear Programs (MILP), which significantly limits the scalability of these techniques to large problems, with the biggest instances solved via FTP being in the order of 800 infosets per player. The biggest crux of these tabular approaches is the need for an explicit representation of the sequential game, which may not be exactly known to players, or could be too big to be stored in memory. For this reason, extremely large games are usually abstracted by bucketing similar states together. The problem with this approach is that abstractions procedures are domain-specific, require extensive domain knowledge, and fail to generalize to new scenarios (see, e.g.,~\cite{gilpin2006competitive,gilpin2007better,sandholm2010state,brown2018superhuman}).

On the other side of the spectrum with respect to tabular equilibrium computation techniques there are Multi-Agent Reinforcement Learning (MARL) algorithms (see \cite{busoniu2008comprehensive,hernandez2019survey} for a comprehensive tractation). These techniques do not require a complete knowledge of the environment and are sample-based by nature, but their application to imperfect-information adversarial team games presents a number of difficulties, such as dealing with private information and representing coordinated strategy spaces. The latter is of crucial importance since we will show that, even in simple settings, it is impossible to reach optimal coordination by exploiting completely decentralized policies, as it is customary in the RL literature.

\paragraph{\textbf{Original Contributions}}
The contribution of the paper is two-fold. First, we study the problem of collecting meaningful histories of play (i.e., trajectory sampling) of the team that can be used, in a subsequent phase, to compute strong coordinated strategies. 
Second, we propose a signal-mediated framework to represent and compute coordinated strategy from a buffer of past experience. 
We address the former problem showing that historical data can be efficiently collected via self-play on an approximated version of the team game, leveraging the notion of its \textit{perfect-recall refinement}, which provides a natural way to perform centralized trajectory sampling when the objective is uncovering effective coordinated behaviors. Moreover, when the team members have symmetric observations, this approach allow us to prove that an optimal team coordinated strategy can be computed in polynomial time, on an equivalent two-player zero-sum game. 
Finally, we propose \emph{signal mediated strategies} (SIMS) as a scalable way to capture coordination without the need for an explicit description of the underlying game. 
Specifically, SIMS represents a coordinated team strategy as the combination of a signaling policy (i.e., a distribution over signals) and one decentralized policy for each team member. First, a signal is sampled from the signaling policy and communicated to the team members. Then, each team member uses the signal to condition the output of a neural network encoding his/her decentralized policy. 
Therefore, in order to approximate an optimal coordinated strategy, team members have to learn from past experiences both the signaling policy and the {\em meaning} associated to each signal (i.e., a suitable parametrisation of the decentralised policies). 
We show that this is possible by testing our framework on a set of coordination games in which previous state-of-the-art multi-agent RL techniques could not reach an optimal coordinated strategy, and on an instance of a simple patrolling game defined over a grid-world.

%% file: text/2.related_work.tex
\section{Related Works}
\label{sec:related_work}
Learning how to coordinate multiple independent agents~\cite{claus1998dynamics, boutilier1999sequential} via Reinforcement Learning requires tackling multiple concurrent challenges, \eg, non-stationarity, alter-exploration and shadowed-equilibria~\cite{matignon2012independent}. There is a rich literature of algorithms proposed for learning cooperative behaviours among independent learners. Most of them are based on heuristics encouraging agents' policies coordination~\cite{bloembergen2010lenient, bowling2001rational, lauer2000algorithm, lauer2004reinforcement, matignon2007hysteretic, matignon2008study, panait2005cooperative}.

Thanks to the recent successes of deep RL in single-agent environments~\cite{Mnih13, silver2016mastering, silver2018general}, MARL is recently experiencing a new wave of interest and some old ideas have been adapted to leverage the power of function approximators~\cite{omidshafiei2017deep, Palmer2018}. Several successful variants of the Actor-Critic framework based on the \emph{centralized training/decentralized execution} paradigm have been proposed~\cite{lowe2017multi, foerster2016learning, foerster2018counterfactual, sukhbaatar2016learning}. These works encourage the emergence of coordination and cooperation, by learning a centralized $Q$-function that exploits additional information available only during training. Other approaches factorize the shared value function into an additive decomposition of the individual values of the agents~\cite{Sunehag2017}, or combine them in a non-linear way~\cite{rashid2018qmix}, enforcing monotonic action-value functions. More recent works, showed the emergence of complex coordinated behaviours across team members in real-time games~\cite{jaderberg2019human, liu2018emergent}, even with a fully independent asynchronous learning, by employing population-based training~\cite{jaderberg2017population}.

 Player's coordination is usually modeled from a game-theoretic perspective via the notion of \textit{correlated equilibrium} (CE)~\cite{aumann1974subjectivity}, where agents make decisions following a recommendation function, \ie, a \emph{correlation device}. Learning a CE of \emph{extensive-form games} (EFG) is a challenging problem as actions spaces grow exponentially in the size of the game tree.
A number of works in the MARL literature address this problem (see, \eg, ~\cite{cigler2011reaching, cigler2013decentralized, greenwald2003correlated, zhang2013coordinating}).
Differently from these works, we are interested in the computation of TMECor~\cite{celli2018computational}.

In our work, we model the correlation device explicitly. By sampling a signal at the beginning of each episode, we show that the team members are capable of learning how to associate a precise meaning to a potentially uninformative signal.
Our approach is closely related to the work by~\citet{chen2019signal}, which proposes a similar approach based on exogenous signals. \citet{chen2019signal} suggest that coordination can be engouraged by  maximizing the mutual information between the signals and the joint policy.

%% file: text/3.preliminaries.tex
\section{Preliminaries}
In this section we provide a brief overview of extensive-form games (see also the textbook by~\citet{shoham2008multiagent}).

An extensive-form games $\game$ is a tree-form model of sequential interactions involving a set of players $\pl$.
A node $v$ of the tree is defined by all the information on the current state of the game. For instance, in a poker game, a node is determined by the history of actions up to that point, and by the hand of each player. The set of actions available to the relevant player at a node $v$ is denoted by $\aspace(v)$.
Leaf nodes are called \emph{terminal nodes}. We denote the set of terminal nodes by $\mathcal{Z}$. Each player $i\in\pl$ as a payoff function $u_i:\mathcal{Z}\to\mathbb{R}$ which specifies her final reward for reaching a certain leaf. 
Exogenous stochasticity is represented via a {\em chance} player (denoted by \chance) which selects actions with a fixed known probability distribution. Given $z\in\Z$, we denote by $\rho_\chance(z)$ the probability with which the chance player plays so as to reach $z$.

Private information is modeled through the notion of \textbf{information states} (a.k.a.  information sets). An information state $s_i$ of player $i$ comprises all nodes of the tree which are indistinguishable to $i$. Taken together, all information states of player $i$ form a partition of the nodes where $i$ has to act. We denote the set of all information states of player $i$ as $\Ss_i$. Given $s\in\Ss_i$, for any pair of nodes $v,w\in s$, nodes $v$ and $w$ must have the same set of available actions.
As is customary in the related literature, we assume {\em perfect recall}, \ie, no player forgets what he/she knew earlier in the game.

In this setting, we distinguish two fundamental paradigms for strategy representation. 
A \textbf{behavioral strategy profile} for player~$i$ is a collection specifying a point in the strategy simplex for each information state in $\Ss_i$. Formally, for any $s\in\Ss_i$, $\pi_i[s]\in\Delta(\aspace(s))$ specifies the probability distribution according to which player $i$ selects an action at $s$.
The second strategy representation is based on the notion of {\em normal-form plan}, which is a vector specifying an action $a_s\in\aspace(s)$ for each information state $s\in\Ss_i$. A \textit{reduced normal-form plan} $p_i$ is a normal-form plan where irrelevant information is removed: it specifies an action only for information states that can be reached following the actions specified by $p_i$ higher up in the game tree. We denote the set of reduced-normal-form plans of $i$ as $P_i$. Given a leaf $z\in\Z$, we denote by $P_i(z)\subseteq P_i$ the set of reduced-normal-form plans in which player $i$ plays so as to reach $z$.
A \textbf{normal-form strategy} for player $i$ is a probability distribution $\mu_i\in\Delta(P_i)$, where $\mu_i[p_i]$ is the probability with which player $i$ will play according to the actions specified by $p_i$.
We say that two strategies of player $i$ are \emph{realization equivalent} if they force the same distribution over the leaves of the game, \ie, the probability of player $i$ playing so as to reach any $z\in\Z$ is the same under both strategies.

%% file: text/4.coordination.tex
\section{Challenges of Team Coordination}

A \textbf{team} is a set of players sharing the same objectives~\cite{VonStengelKoller97,basilico2017team}.
We study games where a team faces an opponent in a zero-sum interaction. In order to simplify the presentation, we describe our results for the case of a team composed by two agents, denoted by \tone~and \ttwo, playing against an opponent \opp. The extension of our framework to the case with multiple team members and opponents is straightforward. We are interested in settings where team members can communicate and agree on a coordinated strategy before the beginning of the game, but are unable to communicate during the playing phase. Two examples of this setting are collusion in poker and collusion during bidding, where \tone~and \ttwo~will share their earning at the end of the game and do not want to be detected, and Bridge, where \tone~and \ttwo~are members of the same team, and the rules of the game do not allow them to communicate during the game.

\subsection{TMECor and Coordinated Strategies}

The most appropriate notion of equilibrium for this setting is the \textit{team-maxmin equilibrium with coordination device} (\textbf{TMECor}) introduced by~\citet{celli2018computational}. 
A powerful, game-theoretic way to think about about coordination is through the notion of \textbf{coordination device}.
Intuitively, before the game starts, team members can identify a set of joint normal-form plans within $P_\tone\times P_\ttwo$.
Then, just before the play, the coordination device draws one of such plans according to a suitable probability distribution $\mu_\team\in\Delta(P_\tone\times P_\ttwo)$, and team members will act as specified in the selected joint plan. A probability distribution over $P_\tone\times P_\ttwo$ is called a \textbf{coordinated strategy}.
A TMECor is a Nash equilibrium (NE) of the game where team members play their best coordinated strategy. 
Let $u_\team:\Z\to\mathbb{R}$ be the shared payoff function of the team. Then, computing a TMECor amounts to solving the following optimization problem:
\begin{equation}\label{eq:tmecor}
\begin{array}{l}\displaystyle
\max_{\mu_\team}\min_{\mu_\opp} \sum_{z\in\Z}\hspace{.4cm}\sum_{\mathclap{\substack{p_\tone\in P_\tone(z)\\ p_\ttwo\in P_\ttwo(z)\\p_\opp\in P_\opp(z)}}} \mu_\team[p_\tone,p_\ttwo]\,\mu_\opp[p_\opp]\,u_\team(z)\\[11mm]
\hspace{.3cm}\textnormal{s.t. }\hspace{.5cm}\mu_\team\in\Delta(P_\tone\times P_\ttwo)\\[2mm]
\hspace{1.3cm}\mu_\opp\in\Delta(P_\opp)
\end{array}
\end{equation}

\noindent By taking the dual of the inner minimization problem, Problem~\eqref{eq:tmecor} can be reformulated a linear programming problem. The main difficulty is then managing the coordinated strategy $\mu_\team$, as its dimension grows exponentially in the size of the game tree. 
In an $\epsilon$-TMECor, neither the team nor the opponent can gain more than $\epsilon$ by deviating from their strategy, assuming that the other does not deviate.

\subsection{Common Pitfalls of Coordination}

By sampling a recommendation from the joint probability distribution $\mu_\team$ the coordination device introduces a correlation between team members' actions that would otherwise be impossible to capture through behavioral strategies.
This is illustrated by the following simple example.

\begin{example}[Coordination game]\label{ex:coordination_game}

The game is played by a team of two players (as usual denoted by\tone~and \ttwo) and an opponent \opp.
Each player of the game has two available actions: left (\lefta) and right (\righta). Players have to select one of the two actions without having observed the choice of the other players.
Team members receive a payoff of $K$ only if they both guess correctly the action taken by the opponent and mimic that action. For example, when the opponent plays \lefta, the team is rewarded $K$ if and only if both \tone~and \ttwo~play \lefta. Otherwise they have payoff equal to $0$.
Team's rewards are depicted in the leaves of the tree in Figure~\ref{fig:coordination_game}.

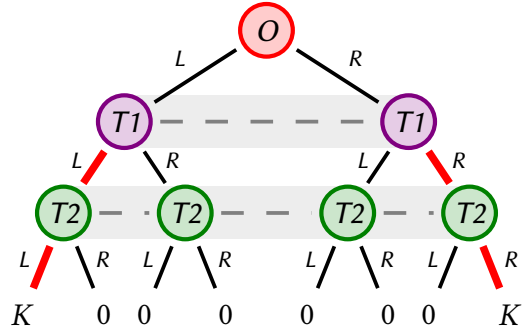
\begin{figure}
	\centering \begin{tikzpicture}[scale=.27]
		\colorlet{darkgreen}{green!50!black};
	    \tikzstyle{treenode} = [circle,black,ultra thick,minimum size=0.7cm,inner sep=0];
	    \tikzstyle{leafn} = [black,minimum size=0.5cm,inner sep=0mm];
	    \tikzstyle{chance} = [fill=white!80!red,draw=red];
	    \tikzstyle{p1} = [fill=white!80!violet,draw=violet];
	    \tikzstyle{p2} = [fill=white!80!darkgreen,draw=darkgreen];

	    \tikzstyle{con} = [draw=black, shorten <=1mm, shorten >=1mm, line width=0.5mm];
		\tikzstyle{highlight} = [line width=1mm,draw=red,shorten <=1mm, shorten >=1mm,];

		\node[treenode,chance] (A) at (0, 0) {\fontsize{12}{0}\opp};

		\fill[fill=black!7!white] (-7, -3.2) rectangle (7, -5.8);
		\node[treenode,p1] (B) at (-7, -4.5) { \fontsize{12}{0}\tone};
		\node[treenode,p1] (C) at (7, -4.5) { \fontsize{12}{0}\tone};

		\fill[fill=black!7!white] (-10, -7.7) rectangle (10.1, -10.3);
		\node[treenode,p2] (D) at (-10, -9) { \fontsize{12}{0}\ttwo};
		\node[treenode,p2] (E) at (-4, -9) { \fontsize{12}{0}\ttwo};
		\node[treenode,p2] (F) at (4, -9) { \fontsize{12}{0}\ttwo};
		\node[treenode,p2] (G) at (10, -9) { \fontsize{12}{0}\ttwo};

		\node[leafn] (H) at (-12, -14) {\fontsize{12}{0}\selectfont {$K$}};
		\node[leafn] (I) at (-8, -14) {\fontsize{12}{0}\selectfont {$0$}};
		\node[leafn] (L) at (-6, -14) {\fontsize{12}{0}\selectfont {$0$}};
		\node[leafn] (M) at (-2, -14) {\fontsize{12}{0}\selectfont {$0$}};
		\node[leafn] (N) at (2, -14) {\fontsize{12}{0}\selectfont {$0$}};
		\node[leafn] (O) at (6, -14) {\fontsize{12}{0}\selectfont {$0$}};
		\node[leafn] (P) at (8, -14) {\fontsize{12}{0}\selectfont {$0$}};
		\node[leafn] (Q) at (12, -14) {\fontsize{12}{0}\selectfont {$K$}};

		\path[con] (A) --node[above left]{\lefta} (B);
		\path[con] (A) --node[above right]{\righta} (C);

		\path[highlight] (B) --node[yshift=1mm, left]{ \lefta} (D);
		\path[con] (B) --node[yshift=1mm, right]{ \righta} (E);

		\path[con] (C) --node[yshift=1mm, left]{ \lefta} (F);
		\path[highlight] (C) --node[yshift=1mm, right]{ \righta} (G);

		\path[highlight] (D) --node[yshift=1mm, left]{ \lefta} (H);
		\path[con] (D) --node[yshift=1mm, right]{ \righta} (I);

		\path[con] (E) --node[yshift=1mm, left]{ \lefta} (L);
		\path[con] (E) --node[yshift=1mm, right]{ \righta} (M);

		\path[con] (F) --node[yshift=1mm, left]{ \lefta} (N);
		\path[con] (F) --node[yshift=1mm, right]{ \righta} (O);

		\path[con] (G) --node[yshift=1mm, left]{ \lefta} (P);
		\path[highlight] (G) --node[yshift=1mm, right]{ \righta} (Q);

		\path[con,gray,dashed,dash pattern=on .3cm off .3cm on .3cm off .3cm] (B) -- (C);
		\path[con,gray,dashed,dash pattern=on .3cm off .3cm on .3cm off .3cm] (D) -- (E);
		\path[con,gray,dashed,dash pattern=on .3cm off .3cm on .3cm off .3cm] (E) -- (F);
		\path[con,gray,dashed,dash pattern=on .3cm off .3cm on .3cm off .3cm] (F) -- (G);
	\end{tikzpicture}
    \caption{Coordination game. Nodes within the same information state are connected through the grey dotted lines. Leaf nodes display the payoff of the team. The payoff of the opponent is $u_\opp(\cdot)=-u_\team(\cdot)$.}
    \label{fig:coordination_game}
\end{figure}

If team members did not have the opportunity of coordinating their strategies, then the best they could do is selecting an action randomly. This corresponds to the Nash equilibrium of the game without coordination, where \tone, \ttwo, and \opp~play a uniform behavioral strategy. This leads to an expected return for the team of $K/4$.
When coordination is possible, team members can skew their joint strategy to play only the reduced-normal-form plans $(\lefta,\lefta)$ and $(\righta,\righta)$ (displayed in red in \cref{fig:coordination_game}). This allows them to avoid playing pairs of actions that would surely result in a $0$ payoff, independently of \opp's action.
The TMECor of the game is reached when \opp~plays with equal probability \lefta~and \righta, and the team plays according to a joint coordinated strategy such that $\mu_\team[(\lefta,\lefta)]=\mu_\team[(\righta,\righta)]=.5$. At the TMECor the team has an expected utility of $K/2$: the team can double its returns by adopting coordinated strategies.
\end{example}

Despite their theoretical superiority with respect to behavioral strategies, coordinated strategies have a major downside in practice: they require an exponential number of bits to be represented. This is because the set of joint reduced-normal-form plans grows exponentially in the size of the game tree. Hence, previous work on the topic largely focuses on providing more manageable representations of the coordinated strategy space (see, \eg, \citet{celli2018computational,Farina&Celli2017}). Here, we take a radically different approach by proposing a model-free framework to approximate coordinated strategies in a RL fashion. 
Classical multi-agent RL algorithms employ decentralized policies, which can be described as behavioral strategies. Therefore, they are unable to attain the optimal coordinated outcome even in simple settings such as the one depicted in the previous example. We show how to reach a middle-ground between decentralized policies (\ie, behavioral strategies) and coordinated strategies. 

%% file: text/5.centralized.tex
\section{Centralized Training for Imperfect-Information Team Games}\label{sec:centralized}

Our framework subdivides the training procedure for approximating a TMECor in two separate phases: in the first phase the game trajectories (\ie, sequences of state-action pairs) are collected in self-play and stored in a buffer $\buffer$. In the second phase, the coordinated team strategy, that will be used at test time, is learned from $\buffer$ via supervised learning.
This approach was already exploited by known algorithms for the two-player, zero-sum setting such as Neural Fictitious Self Play (NFSP) \cite{heinrich2016deep} and Deep-CFR  \cite{brown2018deep}. However, the team coordination setting presents a number of additional challenges: first, unlike in the two-player, zero-sum setting, it is not clear how to collect trajectory samples in such a way to guarantee convergence when two or more team members are coordinating against an opponent. Second, the coordinated strategy to be used at test time must be able to capture coordination without having to represent the exponentially large $\mu_\team$. In this section we propose a solution for the former problem, the latter is discusses in the remainder of the paper (Section~\ref{sec:sims}).

The former problem amounts to populating the buffer $\buffer$ with meaningful trajectories to learn a coordinated strategy. We do that through a centralized training phase in self-play during which we let each team member share with the other its private information. 
We can provide a useful interpretation of this procedure by considering team members $\tone$ and $\ttwo$ as a single meta-player $\team$. In the original game $\game$, $\team$ may have imperfect recall. For example, in Figure~\ref{fig:coordination_game}, $\team$ would not remember his/her first move when choosing its second action. Equivalently, in card-playing game with private cards, $\team$ would have to periodically forget about $\tone$'s cards and regain memory of $\ttwo$'s hand, and vice versa. By letting $\tone$ and $\ttwo$ sharing their private information, we are making sure that $\team$ has perfect recall. This is because information sharing produces finer grained information states. If we denote by $\pref$ the game resulting from this process, we can say that $\pref$ is a \textbf{perfect-recall refinement} of $\game$ for the meta-player $\team$. Following \citet{lanctot2012no}, we define a perfect-recall refinement for $\team$ as follows:
\begin{definition}[Perfect-recall refinement]
\label{def:pr_ref}
Given a game $\game$ with an imperfect-recall player $\team$, $\pref$ is a perfect-recall refinement of $\game$ if $\team$ has perfect recall in $\pref$ and $\game$ is an abstraction of $\pref$, that is if, for any pair of nodes $v,w$ of $\team$ it holds $\aspace(v)\subseteq \aspace^\ast(v)$ and $v,w\in s_\team$, then there exists an information state $s^\ast_\team\in\Ss_\team^\ast$ such that $v,w\in s^\ast_\team$.
\end{definition}
As an illustrative example, the perfect-recall refinement of the game in \cref{fig:coordination_game} is obtained by splitting the information state of $\ttwo$ into two distinct information states: one following action $\lefta$ of $\tone$, and the other following from action $\righta$ of $\tone$.
In a perfect-recall refinement team members share the same observations on the state of the game. Specifically, in $\pref$ team members have imperfect information that can be due only to either partial observability of the actions of the opponent (as it happens in the perfect recall refinement of \cref{fig:coordination_game}), or to private information of the opponent due to a chance moves higher up in the game tree. The key observation is that in $\pref$ either $\tone$ and $\ttwo$ both observe an action (of any player, chance included), or they both do not. We say that $\pref$ has \textbf{symmetric observations} for the team. In this setting, the underlying reason for which the meta-player $\team$ has imperfect recall is the limited observability within the team: that is, $\tone$ not being able to observe every $\ttwo$'s action and vice versa. 

\subsection{Coordination in Games with Team Symmetric Observability}\label{sec:symmetric_observability}

In the class of games in which $\game$ already has symmetric observations for the team, we show that our approach provably yields a TMECor. To show this we need to introduce the notion of \aloss{} recall~\cite{kaneko1995behavior,kline2002minimum}. A player has \aloss{} recall if he/she has perfect recall, or if his/her losses of memory can be traced back to forgetting his/her own actions.
\begin{definition}[Symmetric observability]
A game $\game$ has \emph{symmetric observability} for the team if and only if the meta-player \team{} has \aloss{} recall.
\end{definition}
We observe that a perfect-recall refinement $\pref$ always has symmetric observability for the team, but the converse is not true in general.
A practical example where this condition holds is the game of Goofspiel~\cite{ross1970goofspiel}. In this setting, both team members cannot observe the opponent's move up until the end of each turn, and do not have any private information but the action they just played.

In the following, $\pref$ is always treated as a two-player, zero-sum game between the meta-payer $\team$ and $\opp$. Let $\pi=(\pi_\team,\pi_\opp)$ be an arbitrary behavioral strategy profile of $\pref$.
In order to prove our theoretical results we need the following auxiliary definitions.

\begin{definition}\label{def:mueq}
Two games $\game$ and $\game'$ differing only for their information partitions ($\{\Ss_i\}_{i\in\pl}$ and $\{\Ss'_i\}_{i\in\pl}$, respectively) are $\mu$-equivalent if for any player $i$ and for any normal-form strategy $\mu$ of $i$ in $\game$, there exists a realization equivalent normal-form strategy $\mu'$ in $\game'$, and vice versa. 
\end{definition}

Given a node $v$, let $X_i(v)$ be the set of information state-action pairs of player $i$ on the path from the root of the tree to $v$.
We will make use of the \emph{inflation} operation~\cite{dalkey1953equivalence, kaneko1995behavior, okada1987complete}, which we define as follows:
\begin{definition}[Immediate inflation]\label{def:immediate_inflation}
Let $\Ss_i$ and $\Ss_i'$ be two information partitions of player $i$. We say that $\Ss_i'$ is an immediate inflation of $\Ss_i$ iff there exists $s\in\Ss_i$ and $s',s''\in \Ss_i'$ such that: (i) the set of nodes comprised by $s$ is equal to the set of nodes comprised by $s'$ and $s''$ (i.e., $s=s'\cup s''$), and (ii) for each $v\in s'$ and $w\in s''$ there exists $\bar s\in \Ss_i\cap \Ss_i'$ such that $(\bar s,a)\in X_i(s')$, $(\bar s,b)\in X_i(s'')$ for some actions $a\neq b$.
\end{definition}

\begin{definition}[Inflation]\label{def:inflation}
Given a player $i$, an information partition $\Ss_i'$ is an inflation of $\Ss_i$ iff it is obtained by successive applications of immediate inflation operations to $\Ss_i$.
\end{definition}

When an inflation of $\Ss_i$ has no further immediate inflations, it is
called \emph{complete inflation}.

By leveraging the notion of inflation we can prove the following result, which constitutes a strong motivation for our approach to the computation of a TMECor. 
\begin{restatable}{theorem}{lemmaRealEq}\label{th:real_eq}
Given a game $\game$ with symmetric observability, for any $\pi=(\pi_\team,\pi_\opp)$ which is an NE of $\pref$ there exists a pair of realization equivalent strategies $(\mu_\team,\mu_\opp)$ which is a TMECor of $\game$, and vice versa.
\end{restatable}
\begin{proof}
Given $\game$, let $\team$ be the team meta-player. Formally, $\team$'s information states are such that 
\[
\Ss_\team=\Ss_\tone\cup\Ss_\ttwo.
\]
Then, the set of coordinated strategies of the team $\Delta(P_\tone\times P_\ttwo)$ is equal to the set of normal-form strategies of \team{}, i.e., $\Delta(P_\team)$.
We are left with a two-player, zero-sum game between \opp{} and \team{}. \opp{} has perfect recall and, by Def.~\ref{def:pr_ref}, \team{} has \aloss{} recall. Then, $\game$ is \aloss{}.

\textit{Case \circled{1}: from $\mu_\team$ to $\pi_\team$.}
By Theorem 5.A of \citet{kaneko1995behavior} we have that since the information partition of $\game$ satisfies the \aloss{} condition, then the complete inflation of $\game$ coincides with the perfect-recall refinement $\pref$.
Since the inflation procedure preserves the same $\mu$-equivalence class, we have that $\game$ and $\pref$ are $\mu$-equivalent.
Hence, if we denote by $\mu_\team$ a TMECor of $\game$, by Definition~\ref{def:mueq} there exists a normal-form strategy $\mu^\ast$ of the team meta-player in $\pref$ which is realization equivalent to $\mu_\team$.
By Kuhn's theorem~\cite{kuhn1953extensive}, every normal-form strategy of $\pref$ has an equivalent behavioral strategy: there exists a behavioral strategy $\pi_\team^\ast$ of the team meta-player in $\pref$ which is realization equivalent to $\mu_\team^\ast$, which implies realization equivalence to $\mu_\team$.
By definition of \team{}, $\Delta(P_\tone\times P_\ttwo)=\Delta(P_\team)$. Then, since \opp's information partition in left unchanged going from $\game$ to $\pref$, if $\mu_\team$ is an NE with strategy space $\Delta(P_\tone\times P_\ttwo)$, then $\mu_\team^\ast$ and $\pi_\team^\ast$ are NE of $\pref$.

\textit{Case \circled{2}: from $\pi_\team$ to $\mu_\team$.}
The proof follows the same points of the previous case. 
\end{proof}

Hence, \cref{th:real_eq} justifies the introduction of our centralized training regiment over the perfect-recall refinement of each game, since in many cases this implies sampling trajectories from true equilibrium strategies. 
Then, we have the following key result, which is in striking contrast from with impossibility results by \citet{celli2018computational}.

\begin{restatable}{theorem}{tmecorPoly}\label{thm:poly}
For any game $\game$ with symmetric observability, a TMECor can be computed in polynomial time.
\end{restatable}
\begin{proof}
The result immediately follows from~\cref{th:real_eq} with the following remarks:
\begin{itemize}[leftmargin=*]
\item An NE of a two-player, zero-sum game (i.e.,  $\pref$) can be computed in polynomial time via linear programming by exploiting, for example, the \emph{sequence-form representation} of the game~\cite{koller1996efficient}.

\item The complete inflation of an arbitrary game $\game$ can be computed in polynomial time~\cite[Theorem 3.3]{celli2019personality}.

\item Given $\pi_\team$ in $\pref$ it is possible to compute the reach probability associated to each leaf node $z\in \mathcal{Z}$  in polynomial-time. From there, a realization equivalent normal-form strategy for \team{} can be computed in polynomial time~\cite[Theorem 4]{celli2019learning}. This is a TMECor strategy for the team in $\game$.
\end{itemize}
\end{proof}

Even in games where $\team$ does not have \aloss{} recall, collecting trajectories on a perfect-recall refinement allows team members to populate the buffer of past experience with meaningful trajectories, which are the result of coordinated play in the ideal setting in which they are able to share information. As we will show in the experimental evaluation, this is essential to compute strong coordinated strategies at test time.
The crucial problem of performing strategy mapping between $\pref$ and $\game$ is entirely handled by the SIMS framework (Section~\ref{sec:sims}).

\subsection{Centralized Training: Trajectory Sampling on Perfect-recall Refinement}
\label{subsec:traj_sampling}
A perfect-recall refinement of a game, as defined in Def.~\ref{def:pr_ref}, provides the team players with the ability to observe the actions played by their team members at each step of the game. This simulates the best-case scenario in which agents can communicate during game-play: hence, we can collect meaningful trajectories on the relaxed game $\pref$ using any trajectory sampling algorithm and still exploit the extra information available at training time. We test our framework using NFSP~\cite{heinrich2016deep} and QMIX~\cite{rashid2018qmix} 
(see Appendix for further details).
Going back to the coordination game with horizon $2$ in Figure~\ref{fig:coordination_game}, the team players' observations can be split in two separate information issues: the state of the game and the teammate action. Consider, for example, an episode of the game in which $\tone$ plays action $\lefta$ and $\ttwo$ plays action $\righta$.  Then, the observations of the players will be: $o_{\tone}: \left\{ o: \righta, a_{\ttwo}: \texttt{None} \right\}$ and $o_{\ttwo}: \left\{ o: \righta, a_{\tone}: \lefta \right\}$, where the played actions are $a = \left\{ a_{\tone}: \lefta, a_{\ttwo}: \righta \right\}$ and
$o_{\tone}$, $o_{\ttwo}$ are both $\righta$ because in the original game there is only one information set for each player.

Before storing the collected trajectory into the buffer, players' observations are purged from the extra knowledge of the teammate's action since this information is not available at execution time.
In the example, only $\tone$'s action is purged from $\ttwo$'s observation as for the multistage nature of the original game the only player that does not observe the action of the other is $\ttwo$. Note that in multi-stage games like the coordination game of the example, in order to purge the observation of $\ttwo$ from the information obtained by observing $\tone$'s action is enough to set the observation of $\ttwo$ equal to the one of $\tone$. In the next section, we explain how to use the collected trajectories to learn the team strategy via signal coordination. 

%% file: text/6.sims.tex
\section{SIMS: Signal Mediated Strategies}
\label{sec:sims}

In this section, we focus on the problem of representing team coordinated strategies and, in doing so, we implicitly solve the problem of mapping strategies of $\pref$ back to the original game $\mathcal{G}$. 

As noted by \citet{Farina&Celli2017}, any coordinated strategy $\mu_\team$ can always be represented as the convex combination of a finite set of behavioral strategy profiles $(\pi_\tone,\pi_\ttwo)$ of the team. Therefore, we exploit a set of exogenous signals to condition team members' decentralized policies. Specifically, the decentralized policy followed by each team member at test time is conditioned on a signal which is sampled just before the beginning of the episode. The questions here are: (i) how to properly learn the probability distribution over signals in order to optimally balance different decentralized policies? (ii) how to make sure team members don't just ignore signals?

\begin{figure*}[t]
    \centering
    \includegraphics[width=0.90\textwidth]{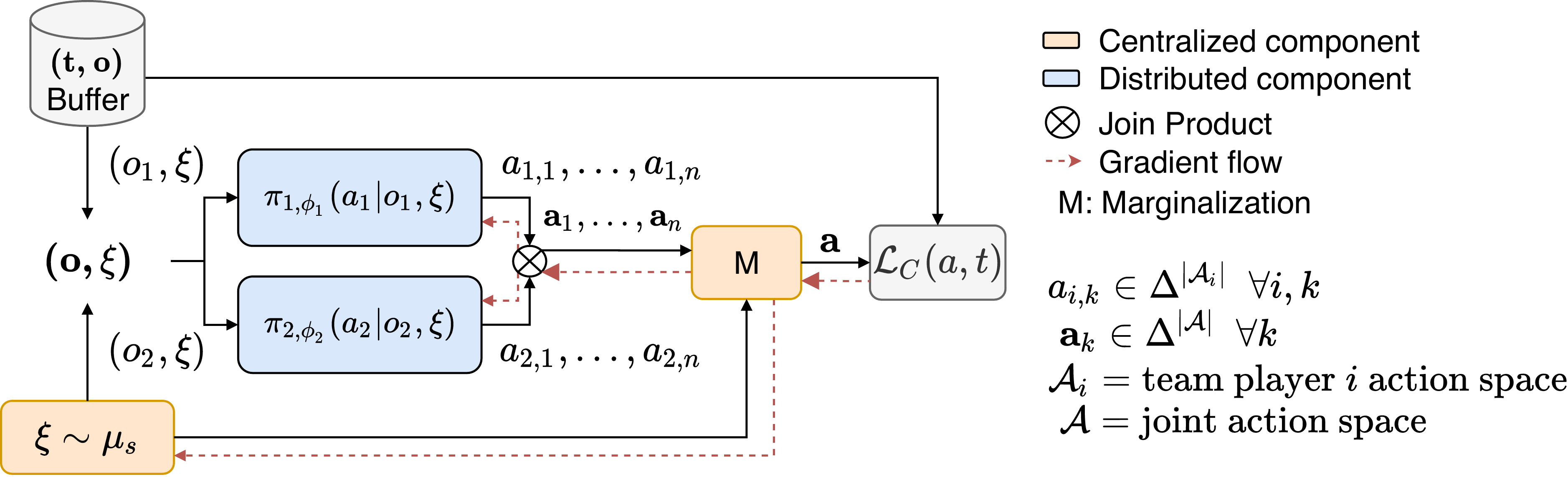}
    \caption{Diagram block of the training of SIMS to learn a coordinated strategy in a game with $2$ team players.}
    \label{fig:sims_framework}
\end{figure*}

Let $\mathcal{M}$ be a memory of trajectories collected from game-play interactions, following the centralized sampling procedure described in Section~\ref{subsec:traj_sampling}. Each sample in $\mathcal{M}$ is a pair $(o,t)$ containing a game observation and its target action. Once the experience buffer is full of meaningful coordinated trajectories, we can try to imitate the team players' behaviour via supervised learning. In practice, coordination is achieved by conditioning the team players' policies on an exogenous signal drawn from a learnable distribution, explicitly implementing a \textit{coordination device}.

\paragraph{\textbf{Signal Mediated Strategies (SIMS)}} Let be $\mathcal{T}$ the set of team players $(\team_1, \team_2, \dots \team_{|\mathcal{T}|})$, with observation space $\mathcal{S}$, and action space $\mathcal{A}_{\team_i}$. 
For simplicity, we assume that the game observation $o$ is the same for all team players and $t$ is a tuple $(t_{\team_1}, ..., t_{\team_{|\mathcal{T}|}})$ that specifies the target action for each team member. In the rest of the section and in Figure~\ref{fig:sims_framework}, for the sake of clarity, we drop the prefix $\team$ from the player notation and denote each player only by its index.

Each team member's policy is defined as $\pi_{i, \phi_i}: \mathcal{S} \times \mathcal{E} \rightarrow \Delta^{|\mathcal{A}_i|}$, where $\mathcal{E}$ is the space of signals and $\phi_i$ are the parameters of a deep neural network that parameterizes the policy.  
Crucially, each policy is conditioned both on the observation $o$ (the state of the game) and on an exogenous signal $\xi\in\mathcal{E}$, sampled at the beginning of the game. The signals are sampled according to a categorical distribution $\mu_s = \textnormal{Cat}(n, \theta)$, where $n$ is the number of available signals and their probabilities are learned from experience, parameterized by $\theta$. 
We sample all the $n$ signals from the distribution $\mu_s$ and compute the action distributions $a_{j, k}$ from $\pi_{j, \phi_{j}}$ for each team member and for each signal $1\leq k \leq n$. 
The marginal action distributions of the team players are then multiplied via joint product to obtain the team action distribution $\mathbf{a}_k$ for each signal $\xi_k$:
\begin{equation}
    \mathbf{a}_k = \prod_{j \in \mathcal{T}} \pi_j(\cdot | o, z_k, \mathcal{\theta}_i) = \prod_{j \in \mathcal{T}} a_{j,k}
\end{equation}
We marginalize the joint action distribution over all the $n$ signals $\mathbf{a} = \sum_{i\leq n} \mu_s[\xi_i]\mathbf{a}_i$, and compute a classification loss $\mathcal{L}_C$ with respect to the target action $t$:
\begin{equation}
     \mathcal{L}_C(\mathbf{a}, \mathcal{\theta}) = \textnormal{CrossEntropy}(\mathbf{a}, t| \mathcal{\theta}).
\end{equation}
We also introduce an additional entropy regularization term $\mathcal{L}_{E, s}=\sum_{t\in\mathcal{T}}H(a_{t,k})$ 
, to enforce pure strategies over the actions probabilities for each distinct signal.
The overall supervised learning loss to be minimized is:
\begin{equation}\label{eq:loss}
    \small
    \mathcal{L}(\mathbf{a}_1,...,\mathbf{a}_n, \mathcal{\theta}) = \mathcal{L}_C + \beta \sum_{k\leq n} \mu_s [\xi_k]\mathcal{L}_{E, s}(\mathbf{a_k})
\end{equation}
where we dropped the arguments of $\mathcal{L}_c$ and $\mathcal{L}_{E,d}$ for clarity, and $\beta$ is the coefficient of the regularization term. A block diagram of the SIMS framework is shown in Figure~\ref{fig:sims_framework}. 

%% file: text/7.experiments_v2.tex
\section{Experimental Evaluation}

\begin{figure*}[ht!]
\centering
\begin{subfigure}[l]{.33\textwidth}
  \centering
  \includegraphics[scale=.25]{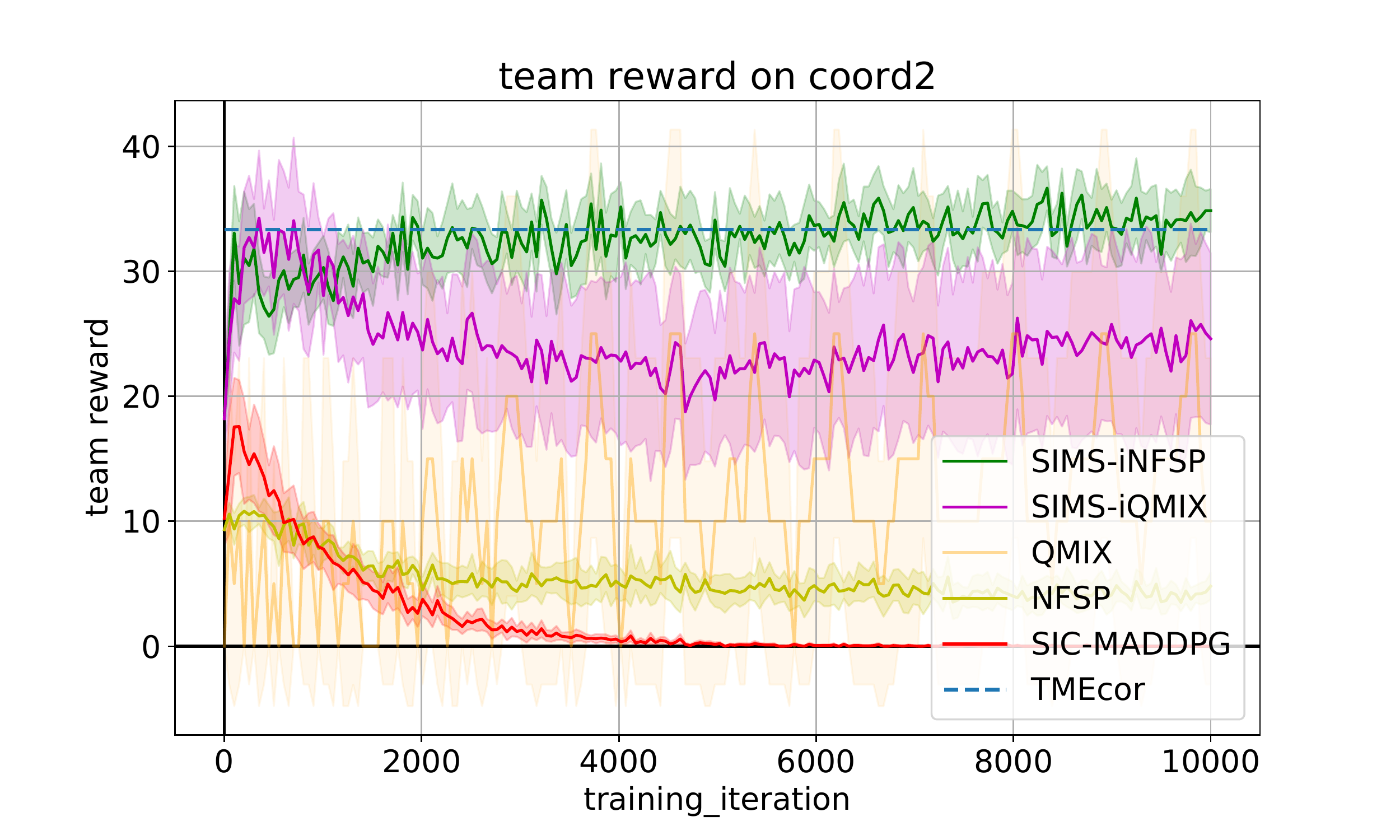}
\end{subfigure}%
\begin{subfigure}[l]{.33\textwidth}
  \centering
  \includegraphics[scale=0.25]{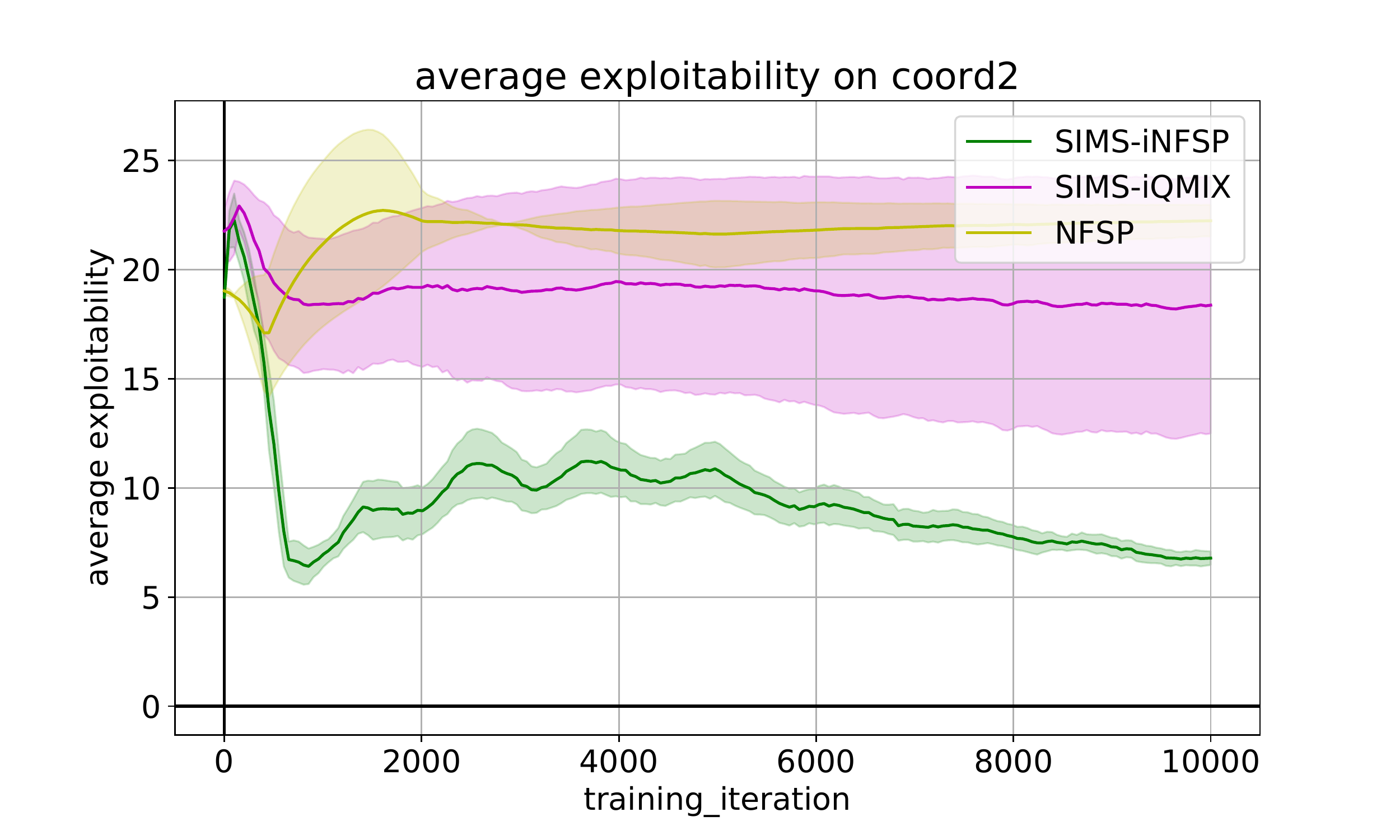}
\end{subfigure}%
\begin{subfigure}[l]{.33\textwidth}
  \centering
  \includegraphics[scale=.25]{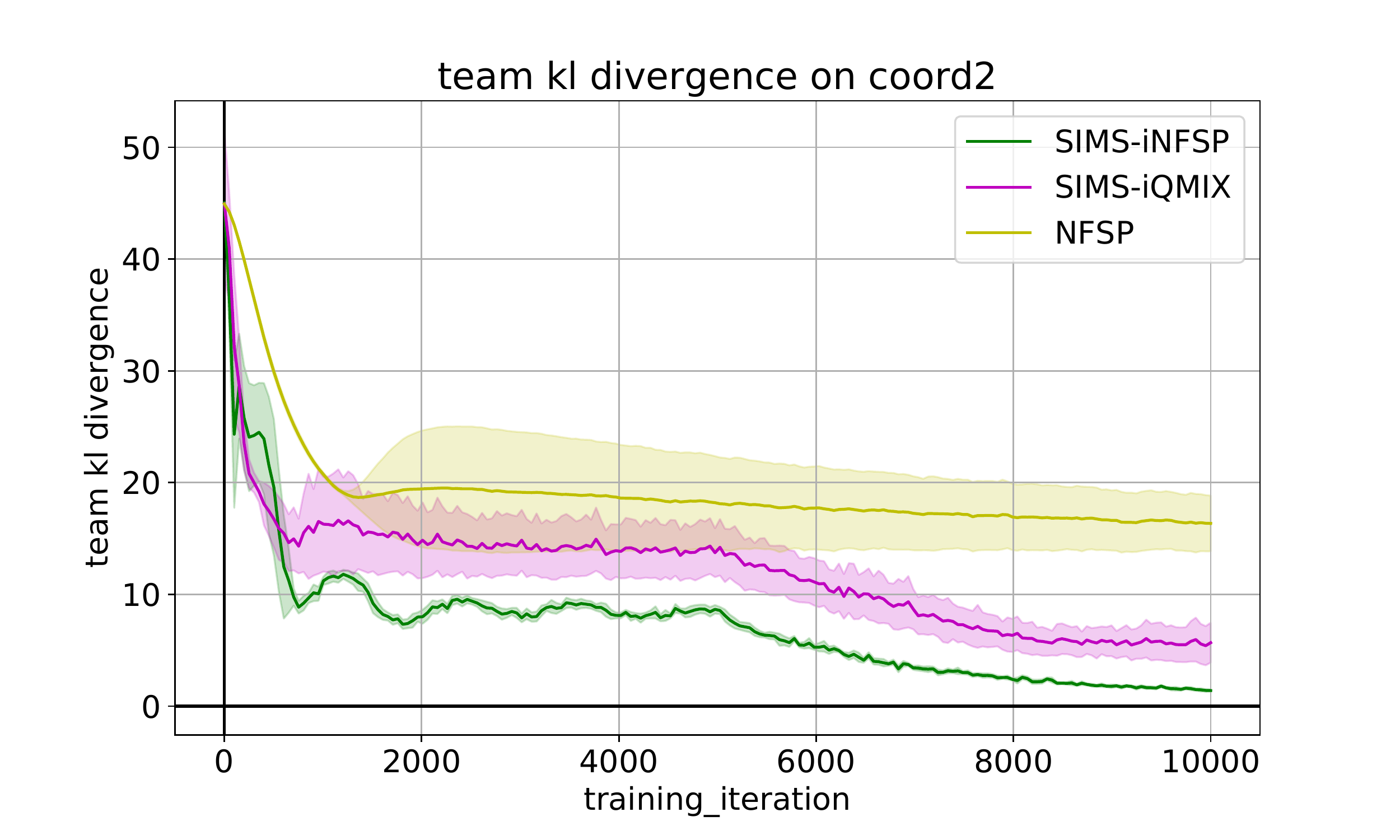}
\end{subfigure}%
\hfill
\begin{subfigure}[l]{.33\textwidth}
  \centering
  \includegraphics[scale=0.25]{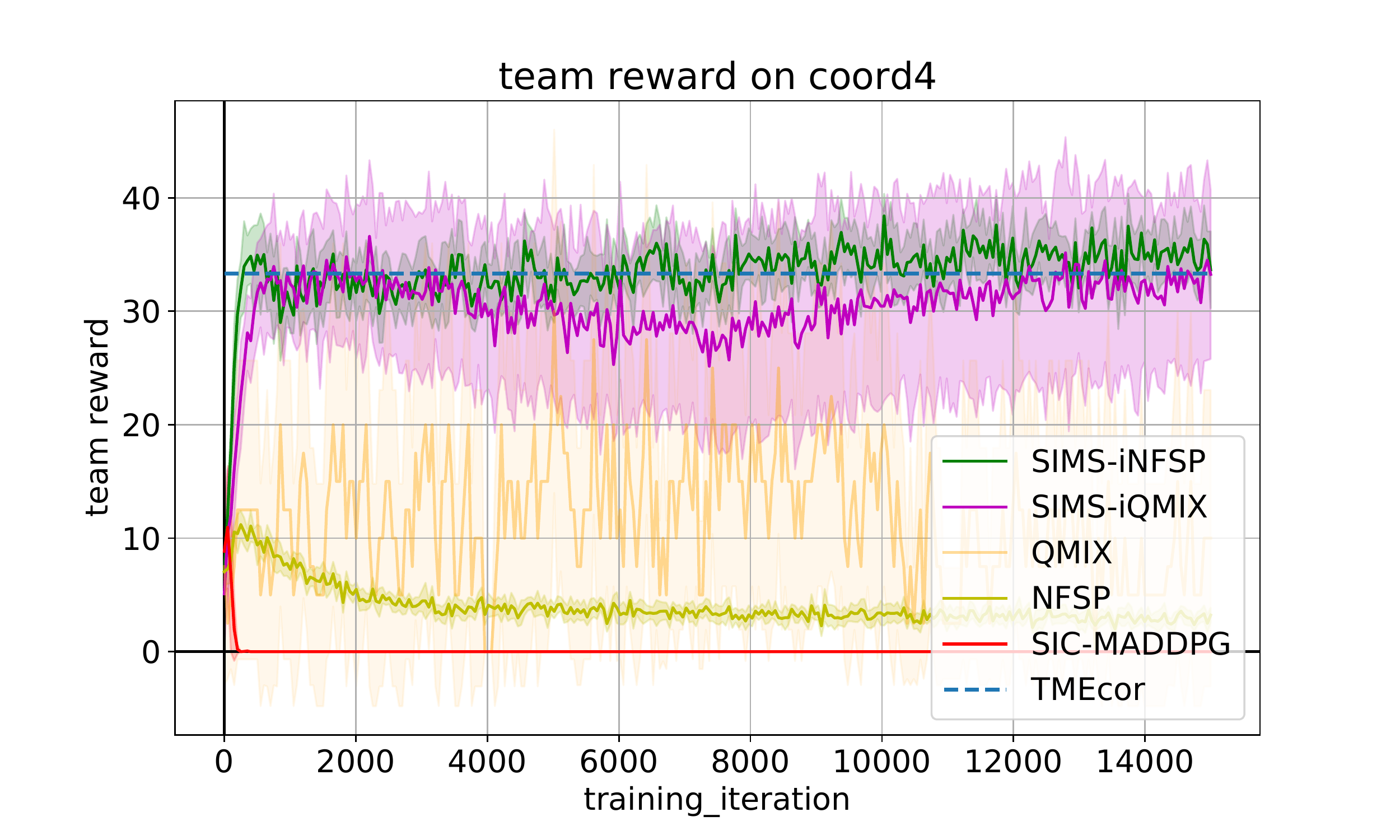}
\end{subfigure}%
\begin{subfigure}[l]{.33\textwidth}
  \centering
  \includegraphics[scale=0.25]{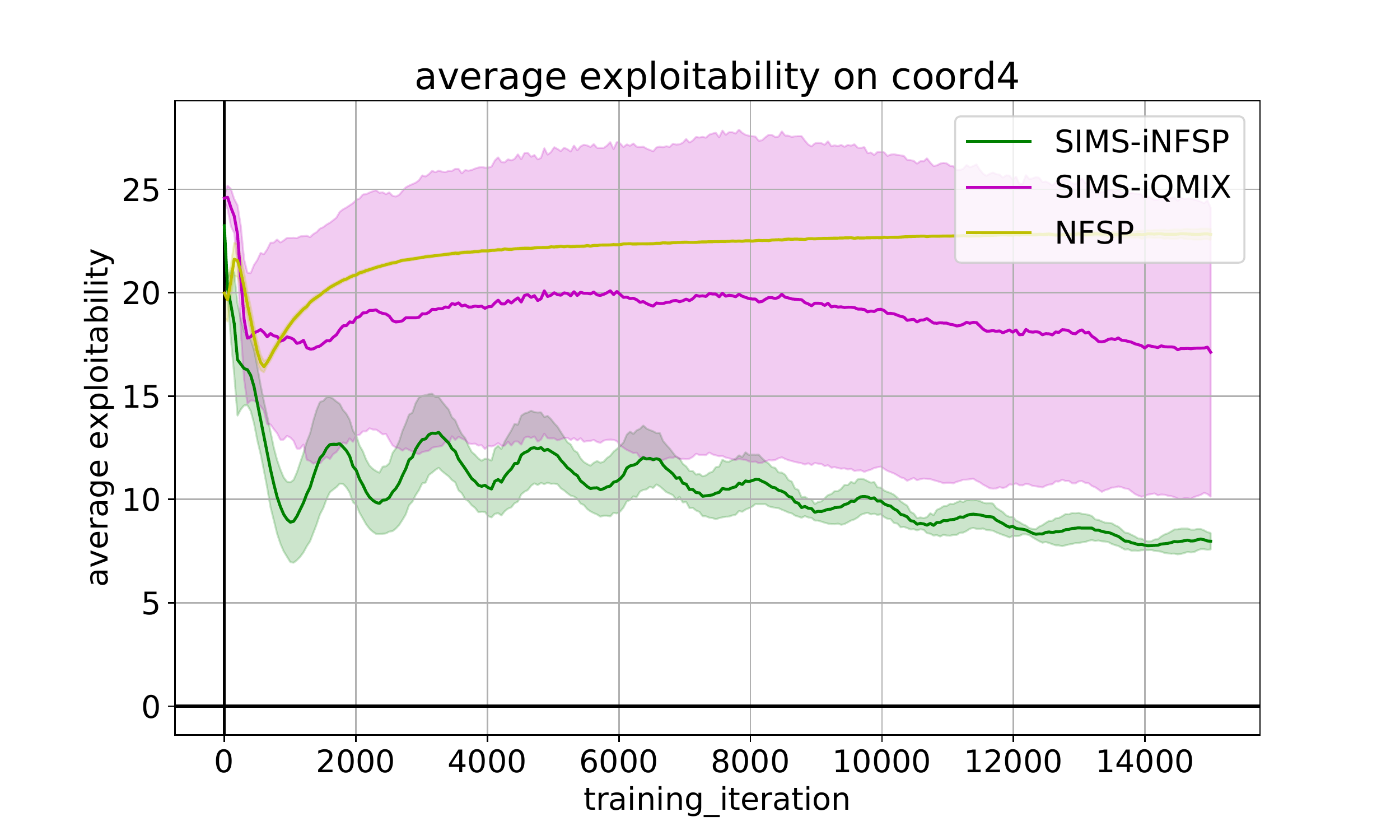}
\end{subfigure}%
\hfill
\begin{subfigure}[l]{.33\textwidth}
  \centering
  \includegraphics[scale=0.25]{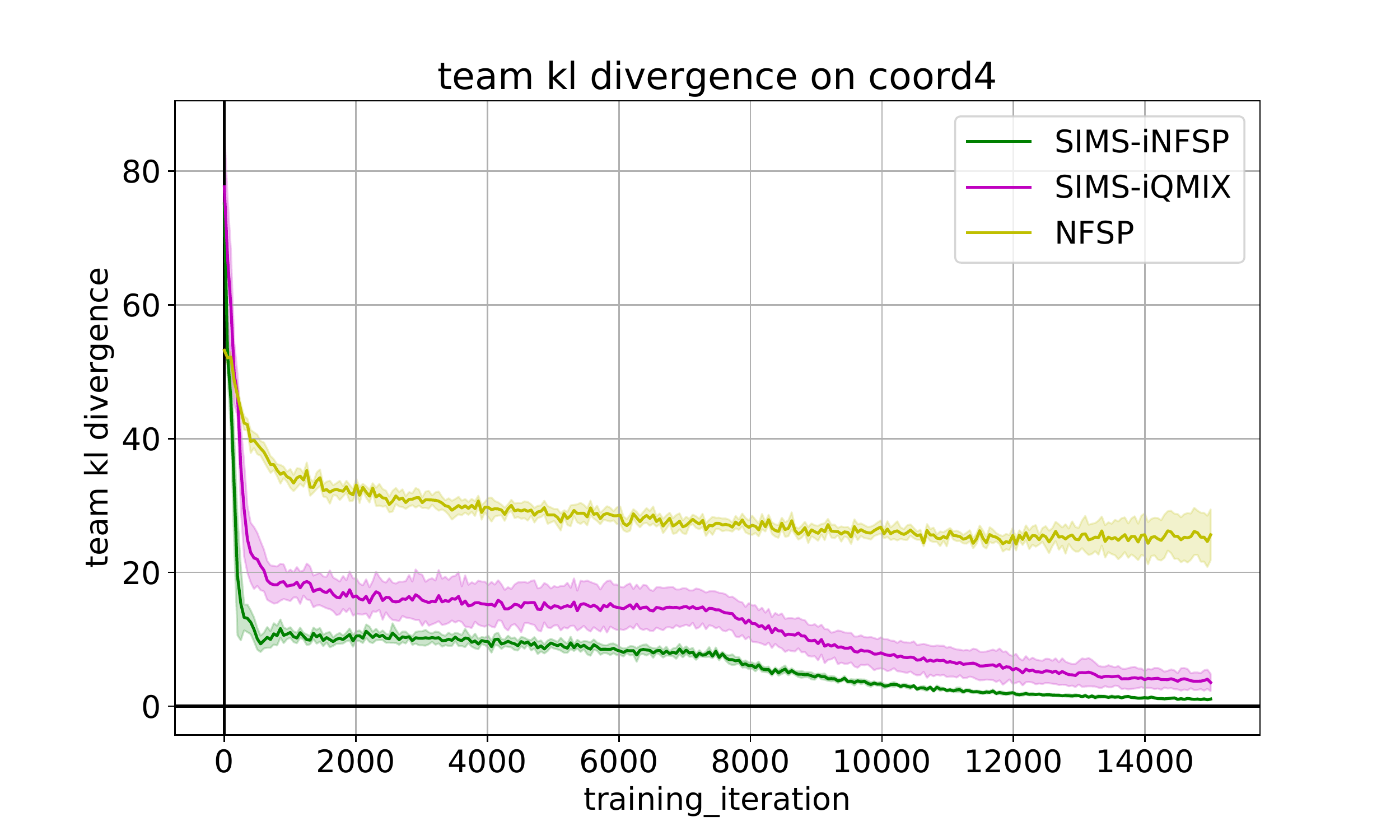}
\end{subfigure}%
\hfill
\caption{\textit{First row:} performance of $\SIMSNFSP$ on a coordination game with horizon 2 (coord-2). \textit{Second row:} performance on coordination game with horizon 4 (coord-4). \textit{Left column:} Average Team reward.
\textit{Center column:} Average exploitability. \textit{Right column:} Kullback-Leibler divergence between the joint average strategy and the optimal TMECor strategy.}
\label{fig:plots_coord_game}
\end{figure*}

In this section, we empirically evaluate our framework against some of the state-of-the-art multi-agent RL algorithms available in the literature. First, we provide some details on our experimental setting, then we provide the main experimental results.

\subsection{Experimental Setting}
\begin{figure}
    \centering
    \includegraphics[scale=0.1]{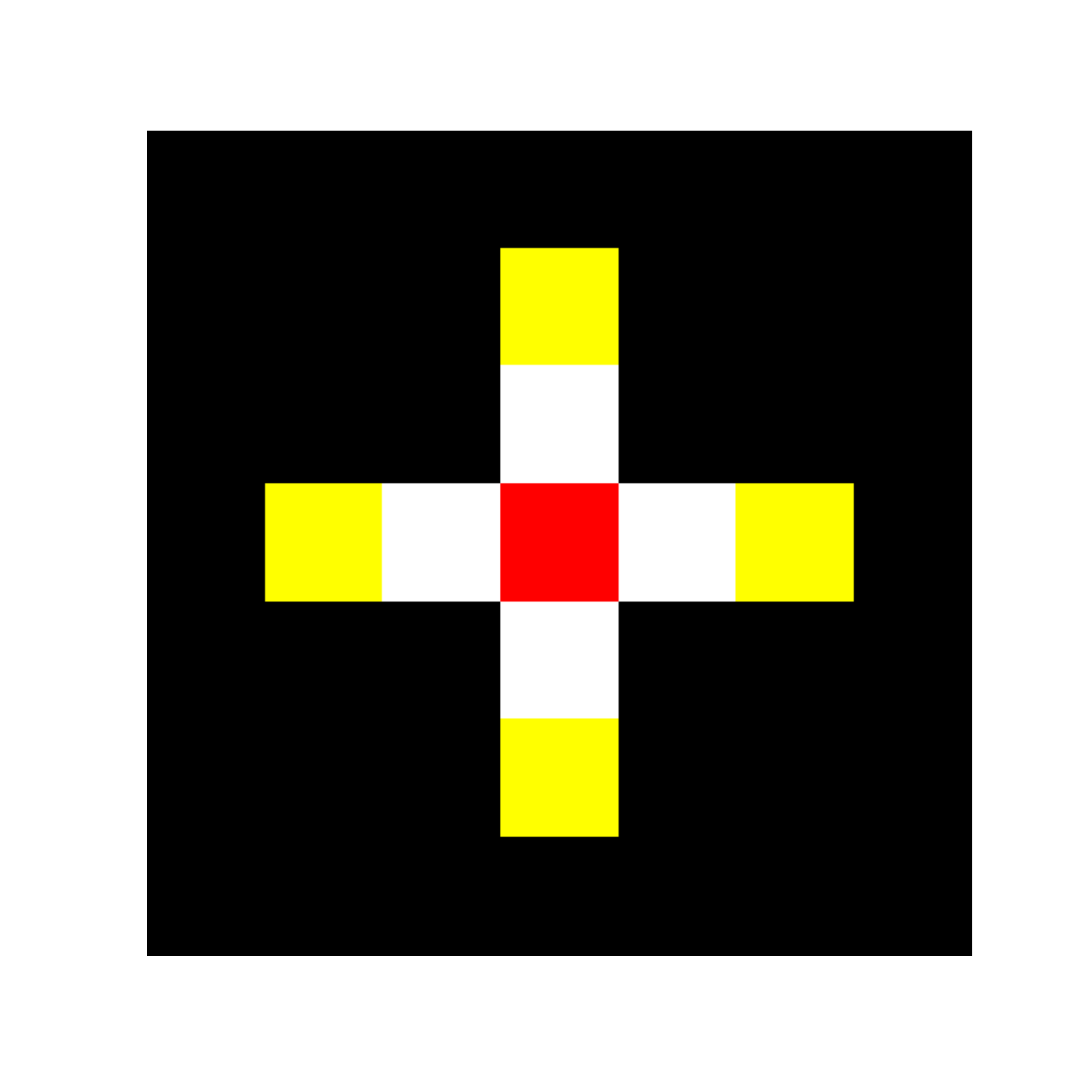}
    \caption{Patrolling game instance. Yellow cells are the facilities that must be defended, the red cell is the starting cell for both defending agents.}
    \label{fig:patrolling}
\end{figure}

We use as benchmarks for our experimental evaluation different instances of the coordination game in Figure~\ref{fig:coordination_game} and an instance of the \textit{patrolling game} shown in Figure~\ref{fig:patrolling}. 

\paragraph{\textbf{Coordination game.}} 
We consider two different instances of the coordination game in Figure~\ref{fig:coordination_game}. Specifically, \emph{coord-$2$} and  \emph{coord-$4$} are coordination games with horizon $2$ and $4$ respectively, \ie, each team member plays once or twice before reaching a terminal node.
In each game, the team receives a positive payoff only if both players guess correctly the action of the opponent and mimic its action. Otherwise, all players have payoff equal to $0$. The final team payoff is split equally among the team members. Note that these game are known to be the worst-case instances in terms of difficulty of coordination~\cite{celli2018computational}, which makes them ideal to understand whether team members are effectively coordinating or not. In particular, we consider the setting in which there is an imbalance in the team’s payoffs, \ie, instead of receiving $K$ for playing both \textit{left} or \textit{right}, team members receive $K$ and $K/2$, respectively. Imbalanced payoffs make the process of learning an optimal coordinated strategy more challenging with respect to the balanced setting. Intuitively, this is because a uniform probability distribution over signals is no longer enough to reach an optimal strategy.
In all the experiments we set $K=100$. We also tested other combinations of imbalanced payoffs, but we omit them since they yield similar results. 

\paragraph{\textbf{Patrolling game.}} We focus on a simple patrolling game defined over a grid-world. The setting is described in Figure~\ref{fig:patrolling}. Both team members play as the defending agents and, starting from the central position in the grid, they have to reach one of the four sites that must be defended. The game evolves synchronously and at every round each agent can either move in one of the four directions or remain on the same cell. After three time instants, the opponent decides which one of the sites to attack (without observing the team members' positions). The defense is considered successful if both agents are located in the exact cell attacked by the opponent, unsuccessful otherwise. In case of a successful defense the team gets as a reward $+1$ and the opponent $-1$, while in case of a successful attack the team gets -1 and the opponent +1. The game is denoted as \textit{patrolling\_$4$\_$3$} to indicate $4$ sites to defend and attacks taking place after $3$ time steps.

\paragraph{\textbf{Baselines.}} We evaluate SIMS in combination with two different algorithms for collecting trajectories, namely $\textbf{\SIMSNFSP}$ and $\textbf{\SIMSQMIX}$. The prefix \textit{``i''} indicates that trajectories have been collected using the relaxed game (\ie, the perfect-recall refinement) and the extra information about the action of the teammate has been discarded when saved into the buffer as described in Section~\ref{subsec:traj_sampling} -- \textit{``i''}~stands for \textit{inflation operator} from Def.~\ref{def:immediate_inflation} and Def.~\ref{def:inflation}.
In order to improve the performances of the trajectory sampling performed by \textbf{\textit{i}NFSP}, we adopt parameter sharing between team members.
We show the necessity of centralized training by sampling the trajectories on the relaxed game and comparing the performance against versions of the \textbf{NFSP} and \textbf{QMIX} algorithms where trajectories have been collected on the original game. Finally, we also perform experiments using \textbf{SIC-MADDPG}~\cite{chen2019signal}, another framework that models the coordination device explicitly. SIC-MADDPG extends the actor-critic framework proposed in~\cite{lowe2017multi} for competitive-cooperative environments by adding signals coordination and ensuring that the signal is taken into account with a mutual information regularizer. MADDPG and hence SIC-MADDPG take advantage of the centralized training by observing extra information such as the state of the game and the actions of other players. In a coordination game, this is equivalent to observing only the actions chosen by other players, as the game can be considered inherently stateless. 
Furthermore, we empirically show that oftentimes SIC-MADDPG and QMIX cannot reach convergence to a TMEcor, even in settings in which \textit{i}NFSP is guaranteed to converge.

\paragraph{\textbf{Implementation details.}}
All the policies and value networks are parameterized by Multi-Layer Perceptron (MLPs) with two fully-connected layers of $128$ units each and \textit{ReLU} activation. The batch size has been set to $128$ and the optimizer used is \textit{Adam}~\cite{kingma2014adam} for all experiments, with a learning rate $lr=10^{-3}$ and default $\beta_1$ and $\beta_2$ parameters. The replay buffers used by all algorithms have a size of $2\cdot 10^4$, while the reservoir buffers in NFSP have a size of $10^5$. The mixing layer used by QMIX has also $128$ units with \textit{ReLU} activation.
In all the considered games, there exists an optimal coordinated strategy employing only two signals. In SIMS, we used a signal space composed of $5$ signals for the coordination games and $4$ signals for the patrolling game. In order to improve stability of the strategy computation, we use a linear scheduling for the entropy regularizer coefficient $\beta$ in Eq.~\ref{eq:loss}. In particular, we keep fixed the parameter at $\beta_{init}$ for the first half of the training, and then linearly increase it up to $\beta_{end}$ for the second half of the training phase. For all our experiments we set $\beta_{init}=0$ and $\beta_{end}=1$. Moreover, we accumulate the gradients of the signals distribution's parameters and perform a back-propagation step every $N_{sig} = 20$ iterations. Both the introduction of the scheduling for $\beta$ and the accumulation of gradients for the signal distribution significantly stabilize the training. 
All the algorithms are evaluated for $100$ episodes every $50$ training iterations by averaging across $10$ different runs. We used PyTorch~\cite{paszke2019pytorch} and the multi-agent environment abstraction from RLlib~\cite{liang2018rllib}.

\subsection{Results}
Now, we discuss the results on the experimental settings described above, showing the success of the SIMS framework in achieving team coordination.

\paragraph{\textbf{Coordination game}} 
In coordination games, decentralized strategies are not expressive enough to describe the optimal coordinated behaviours of the team members. This can be observed by the unsatisfactory performance of NFSP, QMIX and SIC-MADDPG, when compared to the optimal TMECor. 
On the other hand, $\SIMSNFSP$ is able to compute and represent an optimal coordinated strategy for the team. 
As shown in Figure~\ref{fig:plots_coord_game} (Right), the team reward of the joint policy computed through $\SIMSNFSP$ is close to the optimal one reached at the equilibrium. This is further confirmed by~\cref{fig:plots_coord_game} (Center), as the exploitability of the strategy obtained through $\SIMSNFSP$ decreases towards zero. 
Informally, the exploitability of a strategy gives a measure of how much worse that strategy performs versus a best response of the opponent, compared to an equilibrium strategy. 
Finally, the leftmost column of Figure~\ref{fig:plots_coord_game} reports the Kullback-Leibler divergence between the joint policy and the TMECor strategy profile during training. 
The analysis of $\SIMSQMIX$ and its comparison with the other algorithms show two important aspects. On one side it shows how the proposed decentralized training paradigm outperforms the centralized training paradigms of QMIX and SIC-MADDPG, that together with NFSP fail in being able to capture coordination between team members. This happens because, by relaxing the environment, we are giving the possibility to the team members to always avoid playing uncoordinated actions, resulting in an increase of the average reward obtained, and in a decrease in the Kullback-Leibler divergence with the optimal strategy. 
Comparing $\SIMSNFSP$ and $\SIMSQMIX$ further stresses the importance of collecting meaningful trajectories. While the trajectory sampling performed by \textit{i}NFSP has theoretical guarantees of converging to an equilibrium in some specific settings (see Section~\ref{sec:symmetric_observability}), this is not the case for \textit{i}QMIX. This results in high instability of $\SIMSQMIX$'s performances and higher average exploitability, that is a higher distance from the equilibrium.
We also observed that policy gradient approaches struggle to find a pair of strategies with satisfactory performances. We conjecture that this is due to the sparsity of the rewards, a key characteristic of coordination games that inevitably weakens the gradients and prevents the algorithm from learning.

\paragraph{\textbf{Patrolling game}}

\begin{figure}[ht!]
\centering
\begin{subfigure}[l]{.125\textwidth}
  \centering
  \includegraphics[scale=.075]{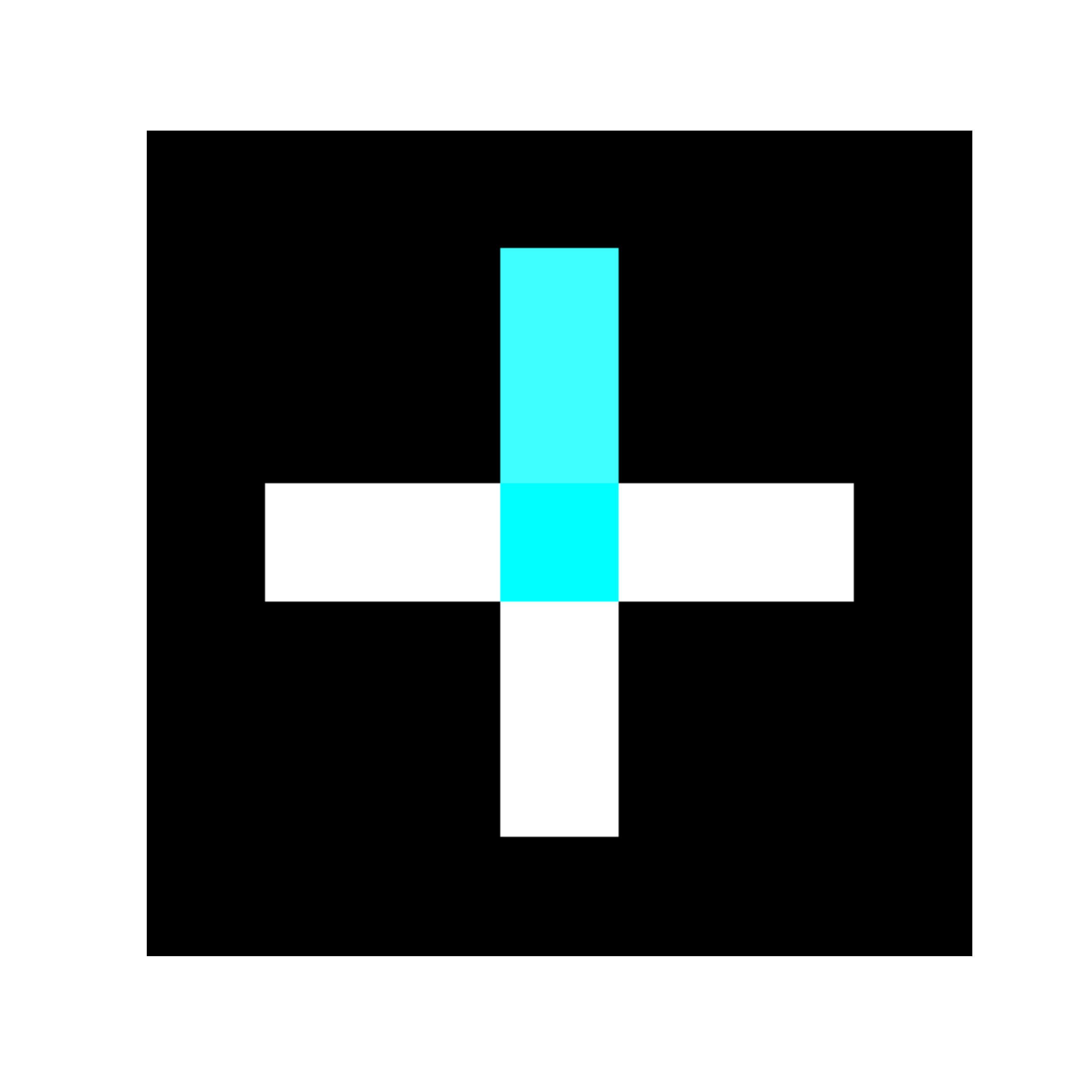}
\end{subfigure}%
\begin{subfigure}[l]{.125\textwidth}
  \centering
  \includegraphics[scale=0.075]{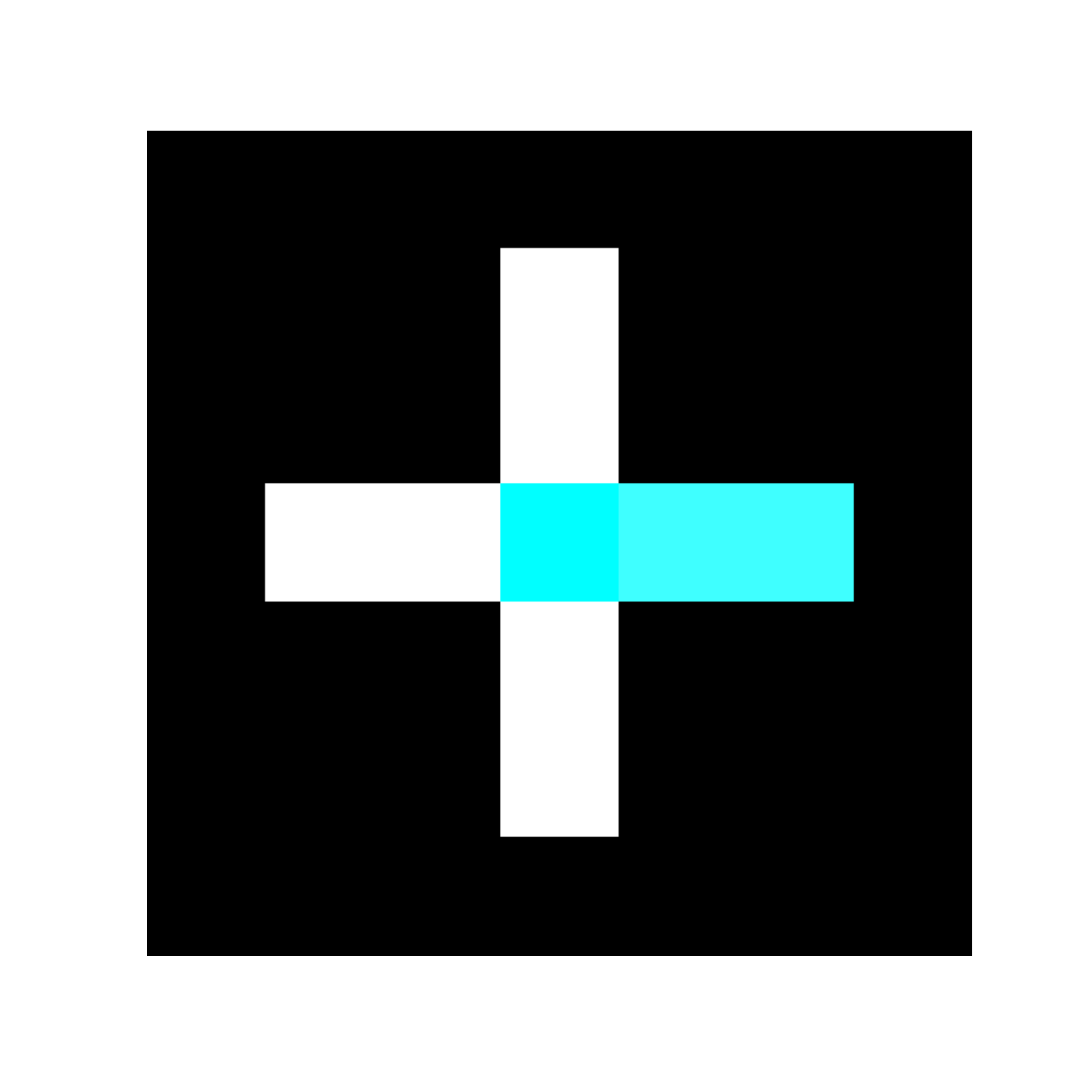}
\end{subfigure}%
\begin{subfigure}[l]{.125\textwidth}
  \centering
  \includegraphics[scale=.075]{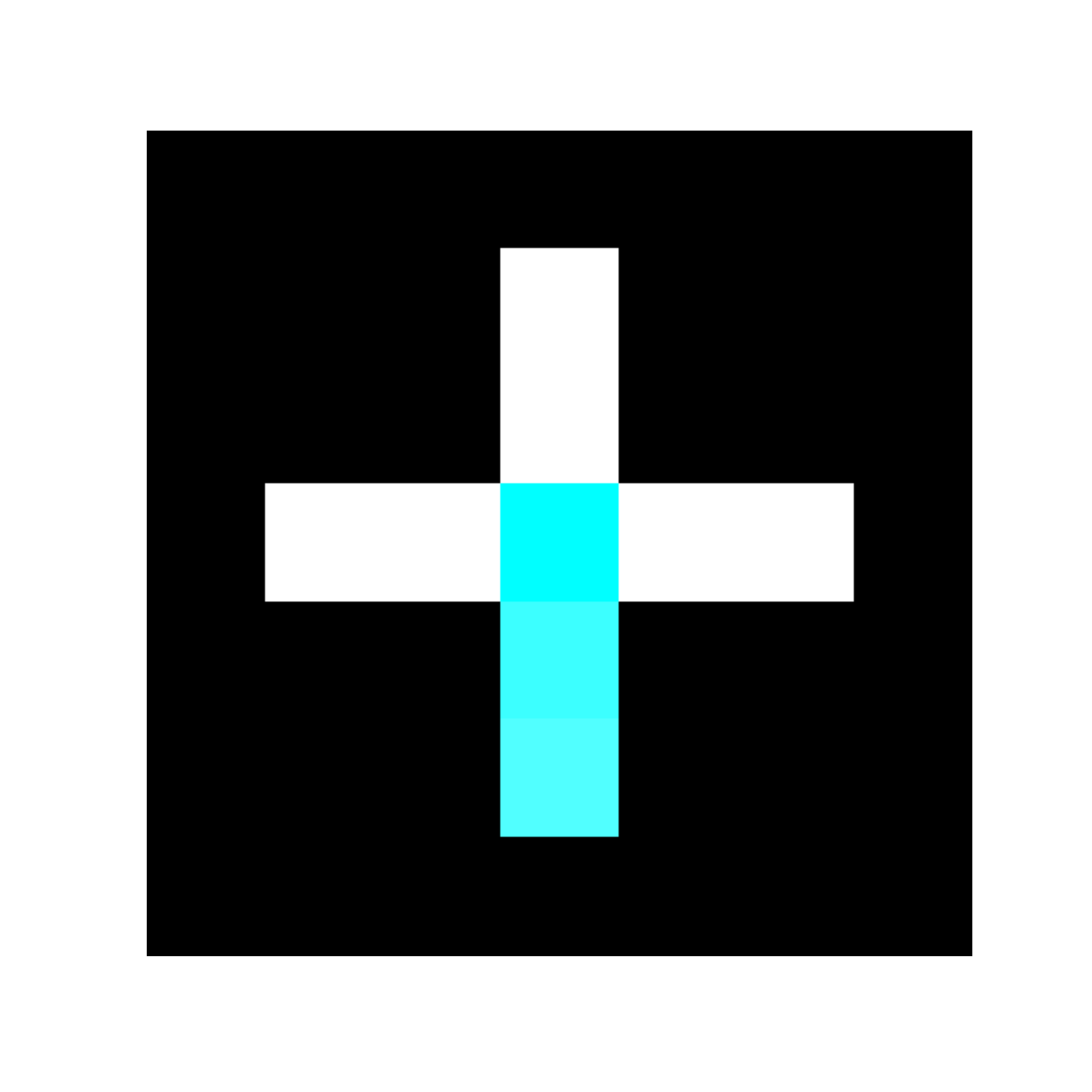}
\end{subfigure}%
\begin{subfigure}[l]{.125\textwidth}
  \centering
  \includegraphics[scale=.075]{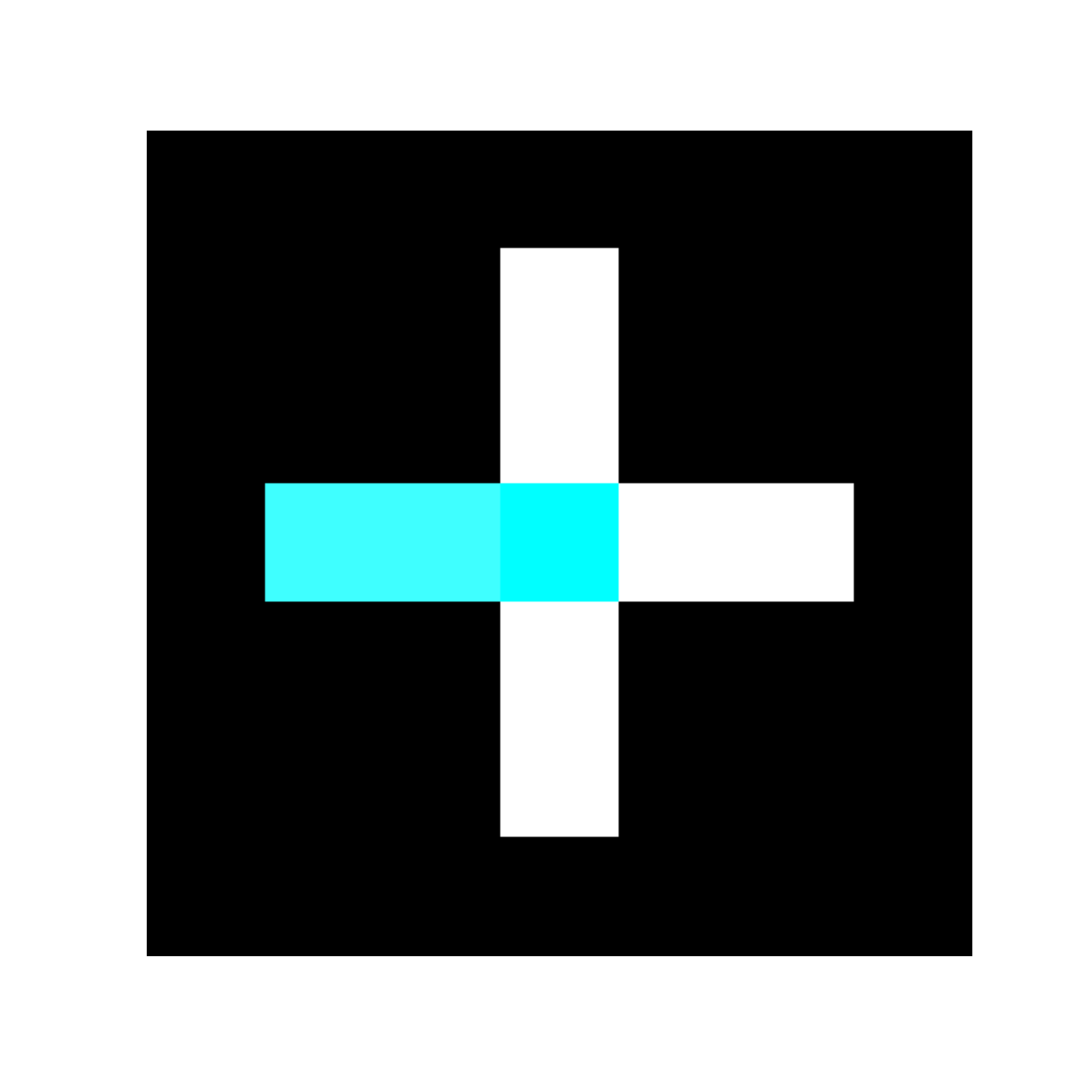}
\end{subfigure}%
\hfill
\begin{subfigure}[l]{.125\textwidth}
  \centering
  \includegraphics[scale=.075]{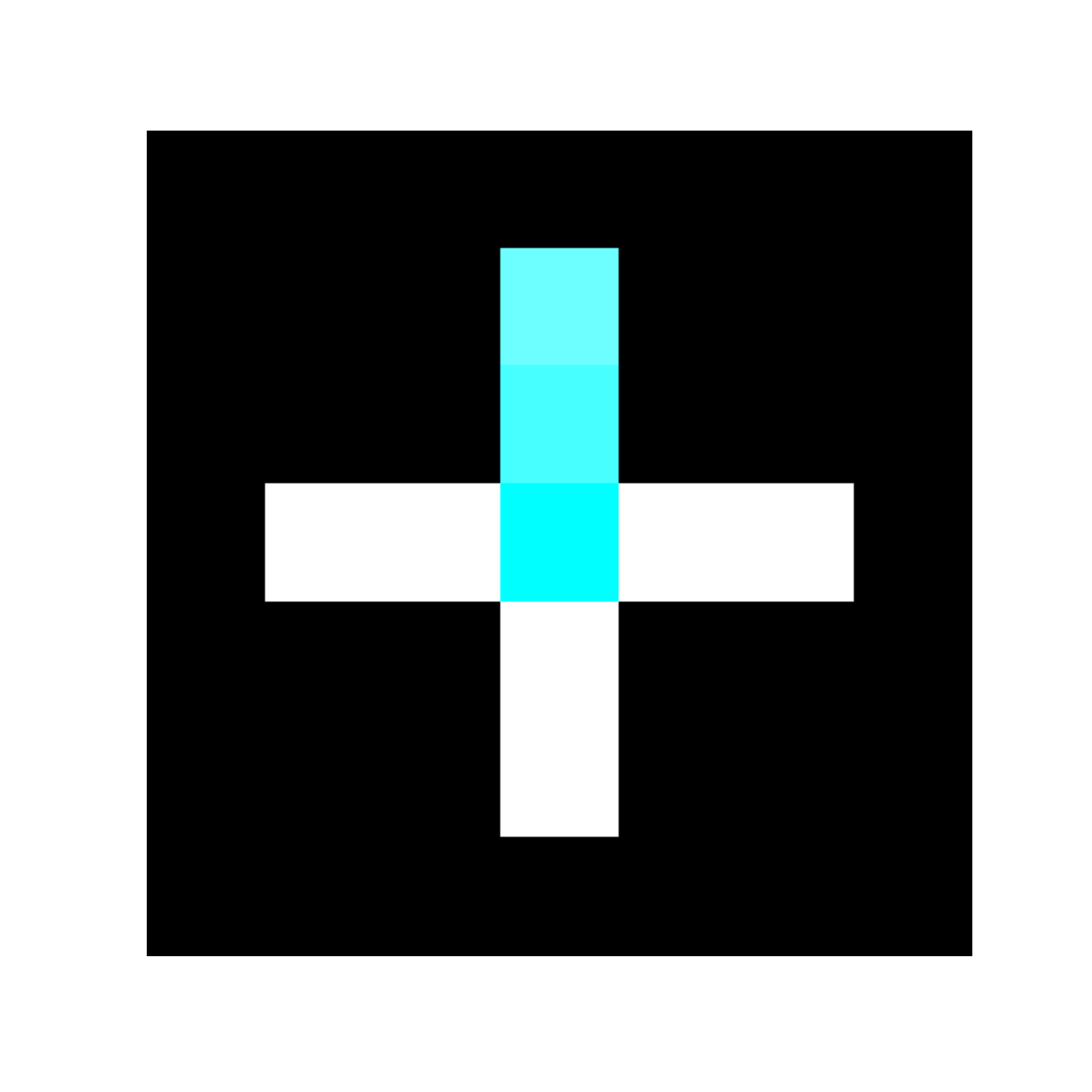}
\end{subfigure}%
\begin{subfigure}[l]{.125\textwidth}
  \centering
  \includegraphics[scale=0.075]{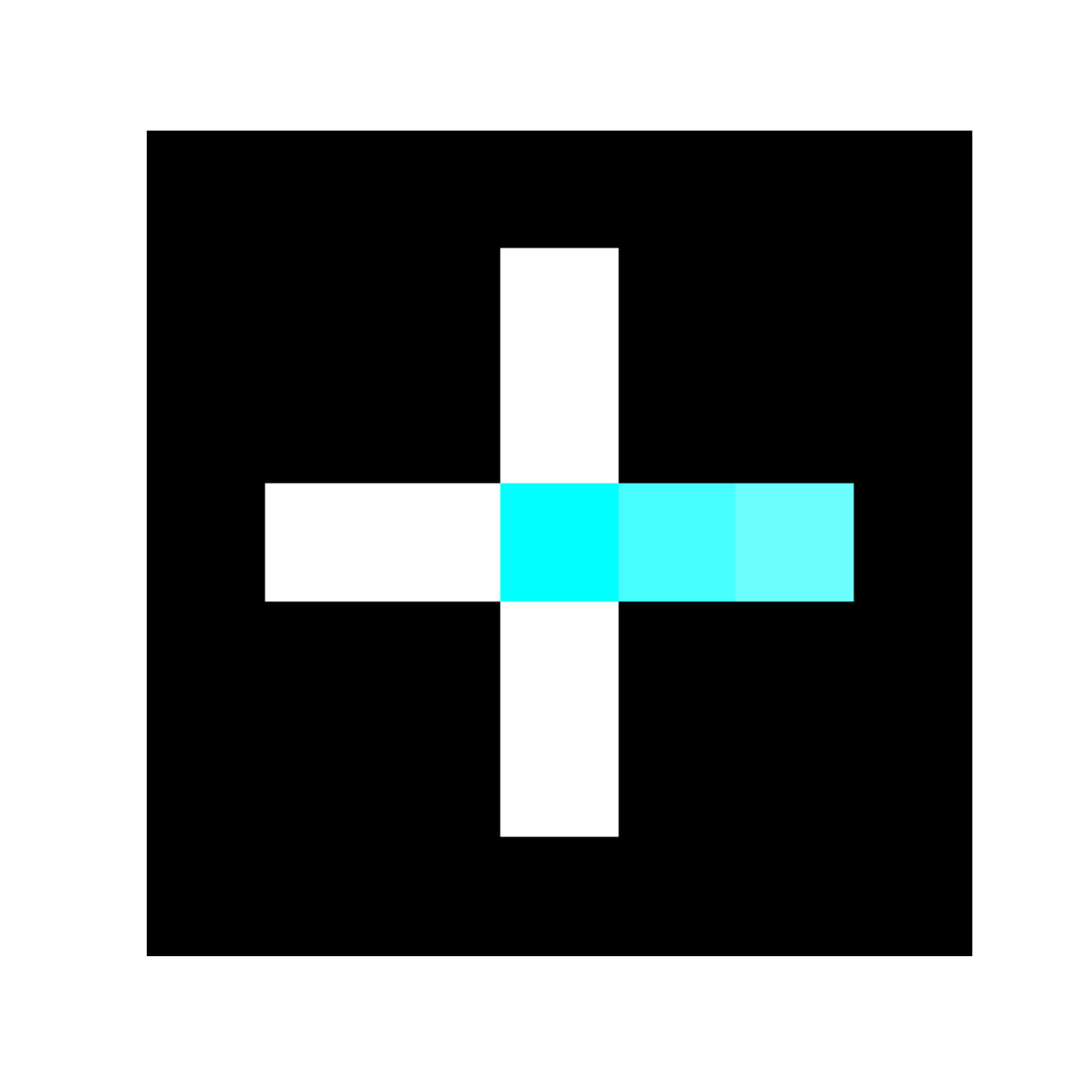}
\end{subfigure}%
\begin{subfigure}[l]{.125\textwidth}
  \centering
  \includegraphics[scale=.075]{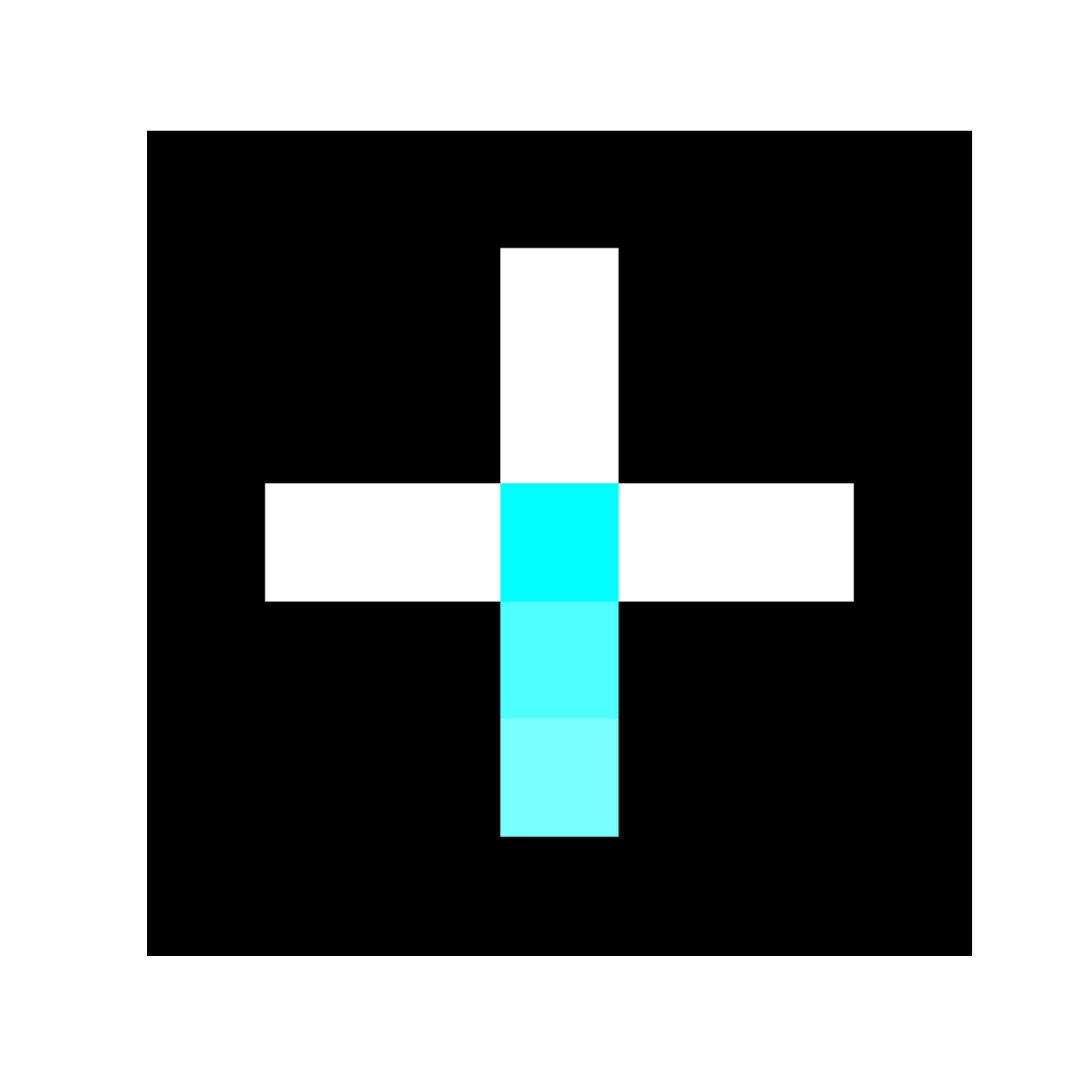}
\end{subfigure}%
\begin{subfigure}[l]{.125\textwidth}
  \centering
  \includegraphics[scale=.075]{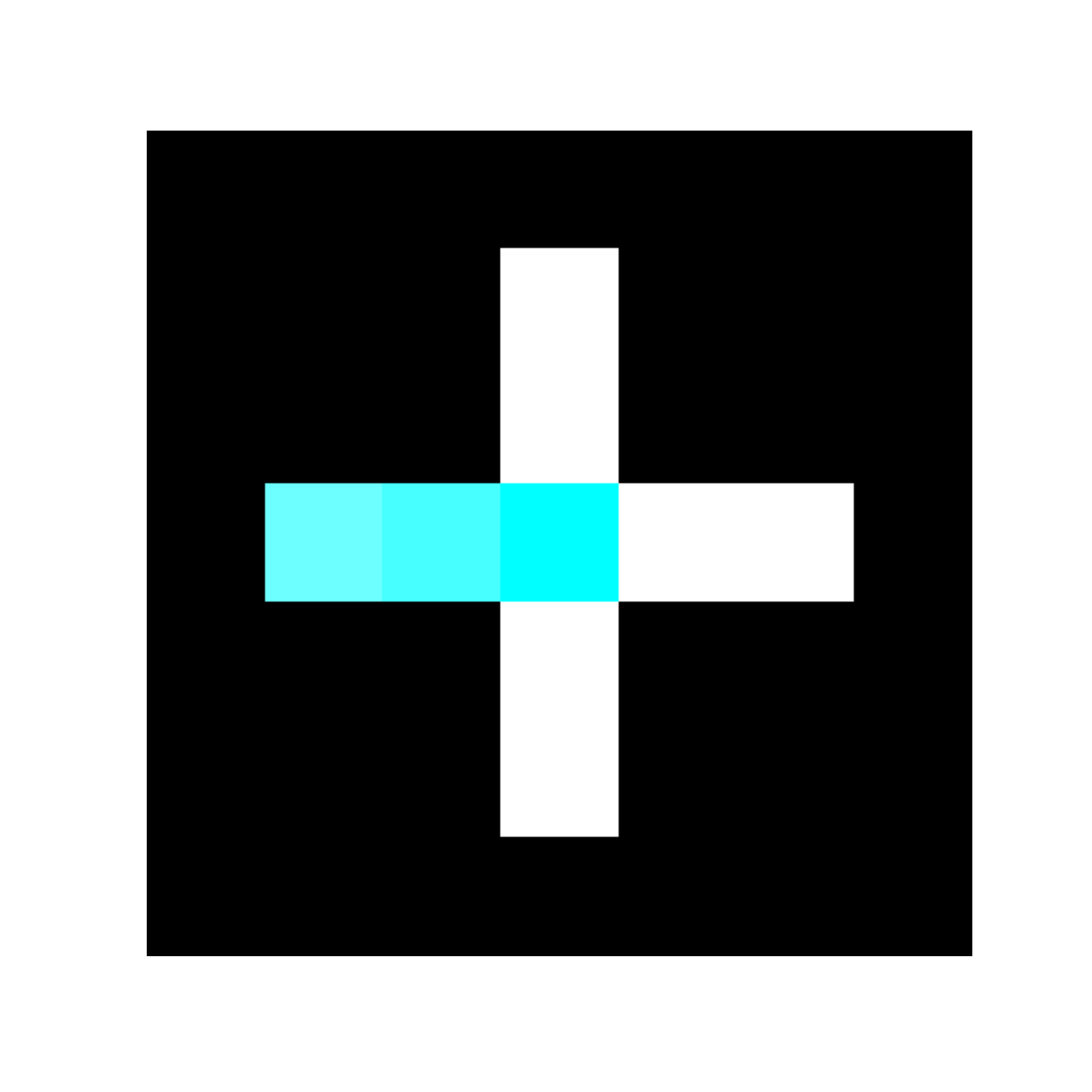}
\end{subfigure}%

\caption{\textit{First row:} heatmaps of player $\tone$ on \textit{patrolling\_4\_3}. \textit{Second row:} heatmaps of player $\ttwo$ on \textit{patrolling\_4\_3}. Every column represents a single signal.}
\label{fig:heatmaps_patr}
\end{figure}

\begin{figure}[ht!]
\centering
\includegraphics[scale=.25]{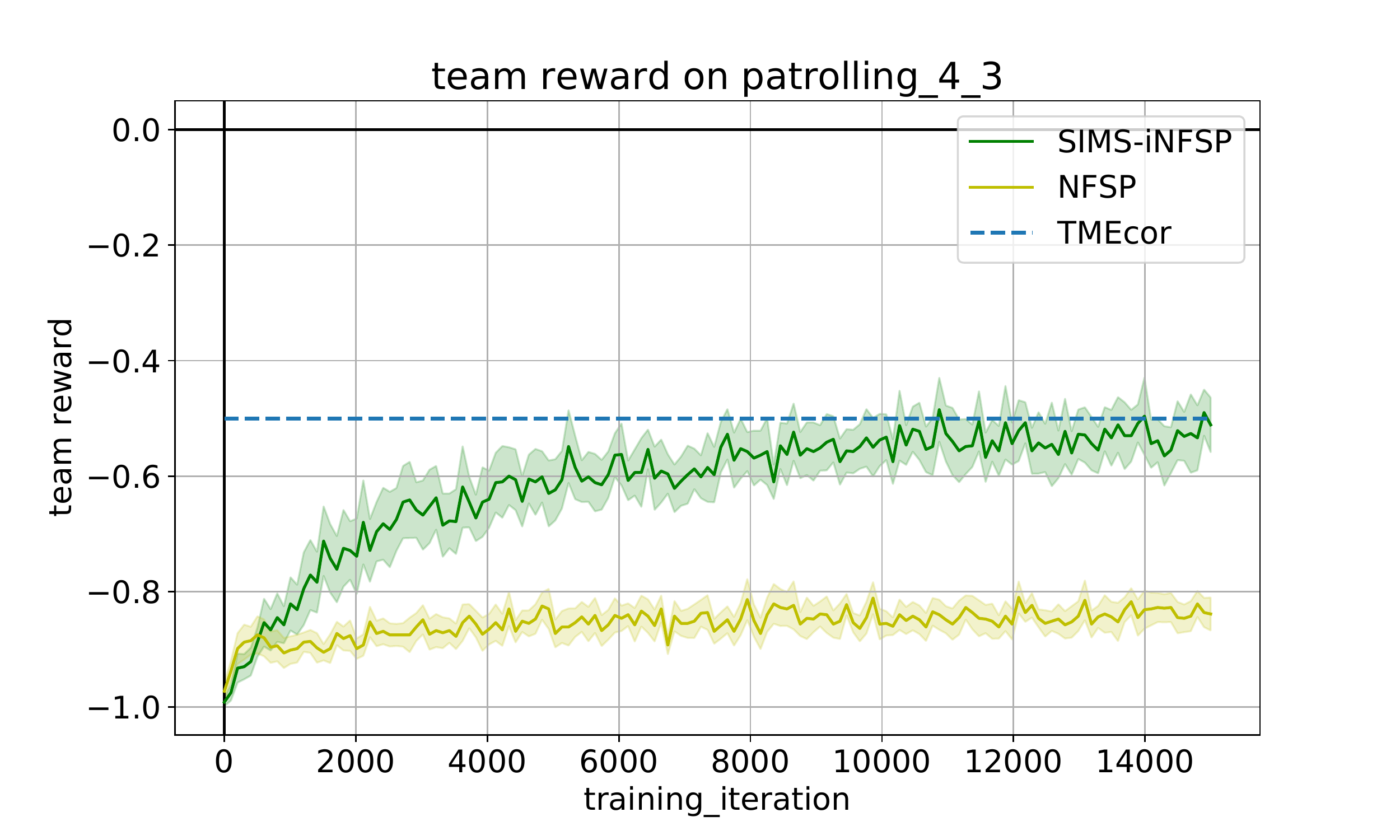}
\caption{Team reward on a patrolling game with $4$ sites to defend and attack after $3$ time steps.}
\label{fig:patrolling_rew}
\end{figure}

The analysis of the patrolling game offers the opportunity to visualize the level of coordination achieved via $\SIMSNFSP$ through the heatmaps that we report in Figure~\ref{fig:heatmaps_patr}.
The heatmaps clearly show that team members learn to associate the same shared meaning to each signal. Specifically, each signal (\ie, each column of Figure~\ref{fig:heatmaps_patr}) is associated with the same site by both team members, resulting in an optimal coordination between them. 
In Figure~\ref{fig:patrolling_rew}, we plot the average reward achieved by $\SIMSNFSP$ against the average reward of NFSP and the reward at the TMEcor. Also in this case, $\SIMSNFSP$ is able to guarantee to team members the optimal reward, resulting in better performances than NFSP. 

%% file: text/8.conclusions.tex
\section{Conclusions and Future Work}\label{sec:conclusions}

In this paper we propose to leverage the notion of perfect-recall refinement of a game to perform centralized trajectory sampling for a team of agents whose goal is to learn a coordinated strategy. Moreover, we introduce a supervised learning framework (SIMS) for computing and representing joint coordinated strategies of a team from a buffer of past experiences.   

We provide a preliminary experimental evaluation which shows promising performance of our framework. Future works will be devoted to further testing the ideas we presented on more challenging benchmarks where the abstraction power of deep RL techniques can be fully appreciated. For example, we plan to test our techniques on predator-prey environments such as Wolpack~\cite{leibo2017multi}. %
We are also planning more extensive tests on different trajectory-sampling algorithms that may be used to populate the buffer of past experiences used by the team.

%% file: supp_content.tex
\appendix
\section{Algorithms for trajectory sampling}\label{sec:appendix_traj_sampling}
Trajectory sampling plays an important role for the convergence of the players' strategies to the equilibrium of the game. We give a brief description of two offline RL algorithms that can be used in our framework to collect trajectories on the relaxed game.

\paragraph{\textbf{Neural Fictitious Self-Play (NFSP)}} \textit{Fictitious Play}~\cite{brown1951} is a game-theoretic self-play algorithm in which players iteratively play their best responses against their opponents’ average strategies. In certain classes of games, e.g. two-player zero-sum~\cite{robinson1951} and many-player potential games~\cite{monderer1996fictitious}, the average strategies are guaranteed to converge to Nash Equilibria (NE), or to approximate $\epsilon$-NE~\cite{leslie2006generalised}, under approximate best responses and perturbed average strategy updates.

The original formulation of Fictitious Play is defined for normal-form games, resulting in exponential complexity for extensive-form games. \citet{heinrich2015fictitious} introduced exact (XFP) and machine learning-based (FPS) versions of the model that are implemented in behavioural strategies. The algorithms are realization equivalent to their normal-form counterpart, inheriting its convergence guarantees to an exact and $\epsilon$-NE, respectively while reducing the complexity from exponential to linear in time and space.

A third variant, used in this work, \textit{Neural Fictitious Self-Play} (NFSP)~\cite{heinrich2016deep}, combines FSP with neural function approximators. In NFSP, agents interact with each others generating datasets of experience in self-play. Each agent collects and stores its experienced transition tuples, $(s_t , a_t , r_{t+1} , s_{t+1})$, in a memory, $\mathcal{M}_{RL}$, while its own best response behaviour, $(s_t , a_t)$, is stored in a separate memory, $\mathcal{M}_{SL}$. Each agent uses off-policy reinforcement learning by training a DQN~\cite{mnih2015human}, $Q(s, a | \theta^{Q})$, to predict action values, from the memory $\mathcal{M}_{RL}$ of its opponents’ behaviour. The agent’s approximate best response strategy is defined as $\beta=\epsilon$-greedy$(Q)$, which selects a random action with probability $\epsilon$ and chooses the action that maximizes the predicted action values otherwise. A separate neural network, $\Pi(s, a | \theta^{\Pi})$, is trained to imitate the agent's own past best response behaviour using supervised classiﬁcation on the data in $\mathcal{M}_{SL}$. This network maps states to action probabilities and deﬁnes the agent’s average strategy, the one that is guaranteed to converge to the equilibrium strategy.

\paragraph{\textbf{QMIX}}
QMIX~\cite{rashid2018qmix} is an offline value-based method that can train decentralised policies in a centralised end-to-end fashion inducing agents' coordination. Learning how to coordinate multiple agents toward cooperative behaviours is hard due to numerous challenges. One can train fully decentralized policies disregarding agents' interactions as in \textit{Independent Q-Learning (IQL)}~\cite{tan1993multi}, but this may not converge because of non-stationary caused by others agents learning and exploration. On the other hand, centralised learning of joint actions can naturally handle coordination problems and avoids non-stationarity, but is hard to scale, as the joint action space grows exponentially in the number of agents. Similarly to \textit{Value Decomposition Networks (VDNs)}~\cite{Sunehag2017}, QMIX lies between IQL and centralised $Q$-Learning. By estimating a joint action-values $Q_{tot}$ as a non-linear combination of per-agent values $Q_{a}$, conditioned only on local observations, QMIX can represent complex centralised action-value functions with a factored representation that scales well in the number of agents. 
QMIX enforces that a global $\argmax$ performed on $Q_{tot}$ yields the same result as a set of individual $argmax$ operations performed on each $Q_a$. This allows each agent to compute decentralized policies by choosing greedy actions with respect to its $Q_a$.
Agents can also be conditioned on their action-observation history to deal with partial observability in the environment by using recurrent neural networks to model their value functions~\cite{hausknecht2015deep}. For more detailed descriptions of the presented algorithms please refer to the original papers.